\documentclass[11pt,cleanfoot]{asme2ej}

\usepackage{helvet}
\usepackage[hyphens]{url}
\usepackage{lineno}
\modulolinenumbers[5]
\usepackage{lipsum} % use blind text
\usepackage{graphicx,ulem}
\usepackage{xspace} 
\xspaceaddexceptions{)} %no space before parenthesis e.g. (\oxygen)
\xspaceaddexceptions{]} %no space before square bracket, e.g. [\Celsius]
\usepackage{textcomp}
\usepackage[]{lmodern}
\usepackage{siunitx} 
\usepackage{tabularx}
\usepackage{floatrow}
\usepackage[font={Large,bf},labelformat=simple]{subfig} 

\usepackage{color}
\usepackage{amsmath}
\usepackage{soul}
\usepackage{booktabs}
\usepackage{paralist}
\usepackage{epstopdf}
\usepackage{setspace}
\usepackage[export]{adjustbox}
\usepackage{textcomp,fancyhdr}
\usepackage[font=normalsize,labelfont=bf]{caption}
\usepackage[plainpages=false]{hyperref}
\hypersetup{final} 
\usepackage[capitalise]{cleveref}

\newcommand{\ds}{decaying\xspace}

\let\subtef\subref
\renewcommand{\subref}[1]{\protect\subtef*{#1}}

\renewcommand{\bar}[1]{\mkern 1.25mu\overline{\mkern-1.25mu#1\mkern-1.25mu}\mkern 1.25mu}

\crefmultiformat{figure}%
{\edef\crefstripprefixinfo{#1}Figs.~#2#1#3}%
{ and~(#2\crefstripprefix{\crefstripprefixinfo}{#1}#3}%
{, #2\crefstripprefix{\crefstripprefixinfo}{#1}#3}%
{, and~#2\crefstripprefix{\crefstripprefixinfo}{#1}#3}

\title{The influence of strut waviness on the tensile response of lattice materials}

\author{Philipp E. Seiler
  \affiliation{
    Department of Engineering\\
    University of Cambridge\\
    Cambridge CB2 1PZ,
    United Kingdom\\
    Email: pes34@cam.ac.uk
    }
}

\author{Kan Li
  \affiliation{
    Department of Engineering\\
    University of Cambridge\\
    Cambridge CB2 1PZ,
    United Kingdom\\
    Email: kl513@cam.ac.uk
    }
  }

  \author{Vikram S. Deshpande
  \affiliation{
    Department of Engineering\\
    University of Cambridge\\
    Cambridge CB2 1PZ,
    United Kingdom\\
    Email: vsd20@cam.ac.uk
    }
}

\author{Norman A. Fleck
  \affiliation{
    Department of Engineering\\
    University of Cambridge\\
    Cambridge CB2 1PZ,
    United Kingdom\\
    Email: naf1@eng.cam.ac.uk
    }
}

\begin{document}

\maketitle    
\doublespacing

\begin{abstract} {\it Recent advances in additive manufacturing
    methods make it possible, for the first time, to manufacture
    complex micro-architectured solids that achieve desired stress
    versus strain responses. Here, we report experimental measurements
    and associated finite element (FE) calculations on the effect of
    strut shape upon the tensile response of two-dimensional (2D)
    lattices made from low-carbon steel sheets. Two lattice topologies
    are considered: (i) a stretching-dominated triangular lattice and
    (ii) a bending-dominated hexagonal lattice. It is found that strut
    waviness can enhance the ductility of each lattice, particularly
    for bending-dominated hexagonal lattices. Manufacturing
    imperfections such as undercuts have a small effect on the
    ductility of the lattices but can significantly reduce the
    ultimate tensile strength. FE simulations provide additional
    insight into these observations and are used to construct design
    maps to aid the design of lattices with specified strength and
    ductility.  }
\end{abstract}

\section{Introduction}
Recent advances in manufacturing
methods~\cite{Tumbleston2015,Vyatskikh2018} have facilitated the
manufacture of lattice materials with complex topologies over a wide
range of length scale. This class of materials is used in a large
variety of engineering applications, e.\,g. tower structures in civil
engineering, the cores of lightweight sandwich panels, and microscopic
mechanical filters~\cite{Phani2017}. In the present study, we consider
the tensile behaviour of 2-dimensional (2D) triangular and hexagonal
lattices. This choice of lattice topology is motivated by the broad
range of their mechanical properties: stretching-dominated triangular
lattices have high in-plane stiffness and strength whereas hexagonal
lattices are compliant, bending-dominated
structures~\cite{Hutchinson2006,Tankasala2017}. Wavy struts of
sinusoidal shape are introduced in order to modify the macroscopic
stress versus strain response of each lattice.

\subsection{Tensile deformation of lattice materials}
\label{sec:tensil_deformation_tria_hex_lattices}
Two-dimensional (2D) lattices of triangular and hexagonal geometry
comprise struts of node-to-node length $\ell$ and in-plane strut
thickness $t$. The macroscopic properties of these lattices scale with
their relative density $\bar\rho$, as defined by the ratio of volume
occupied by solid material to the total volume of the lattice. For
straight slender struts, the relative density $\bar\rho$ scales
linearly with the stockiness $t/\ell$ of each strut according to
\begin{equation}
  \label{eq:relative_dens_straight}
  \bar{\rho}=A\frac{t}{\ell}
\end{equation}
where the coefficient $A$ depends upon the architecture of the lattice
($A$ equals $2\sqrt{3}$ for triangular lattices and $2/\sqrt{3}$ for
hexagonal lattices~\cite{Gibson1999}). It is evident that, for the same
value of $t/\ell$, triangular lattices have a greater density than
to hexagonal lattices. Note that \cref{eq:relative_dens_straight} is
an approximation as it neglects the volume of the nodes of the lattice
but it suffices for low relative densities (typically $\bar\rho<0.2$).

Consider a lattice made from elastic, perfectly plastic cell walls of
elastic modulus $E_\mathrm{s}$ and yield strength
$\sigma_\mathrm{YS}$. Then, the macroscopic in-plane modulus $E$ and
yield strength $\sigma_\mathrm{Y}$ of an infinite periodic lattice
scale with $\bar\rho$ via the relations \cite{Gibson1999}
\begin{equation}
  \label{eq:scaling_law}
  E=C\bar{\rho}^cE_\mathrm{s}\hspace{0.35cm}\text{and}\hspace{0.35cm}\sigma_\mathrm{Y}=D\bar{\rho}^d\sigma_\mathrm{YS}
\end{equation}
where $C=1/3$, $c=1$, $D=1/3$, and $d=1$ for the triangular lattice
and $C=3/2$, $c=3$, $D=1/2$, and $d=2$ for the hexagonal
lattice~\cite{Gibson1999,Fleck2007}.

While Eqs.~(\ref{eq:scaling_law}) characterise the small strain
response, the response under finite deformations is more complex. For
example, the uniaxial tensile response of an elastoplastic hexagonal
lattice exhibits four regimes, as discussed by Tankasala et
al.~\cite{Tankasala2017} and Ronan et al.~\cite{Ronan2016}. The
sequence of deformation modes with increasing applied macroscopic
strain are: (i) elastic bending of struts, (ii) plastic bending of
struts (iii) elastic stretching of struts as the inclined struts
rotate towards loading direction, and finally (iv) plastic stretching
of the struts aligned with the loading direction. In contrast,
triangular lattices deform by stretching at low levels of applied
strain and exhibit three regimes: (i) elastic stretching of struts,
(ii) plastic stretching and rotation of the struts towards the
direction of macroscopic straining, and (iii) a final regime involving
stretching of struts that are aligned with the loading
direction~\cite{Tankasala2017}. Failure may intervene during any of
these regimes depending upon the properties of the cell wall material.

\subsection{Classes of lattices}
\label{sec:classes_lattices}
Typically, lattices are bending-dominated (hexagonal) or stretching-dominated
(triangular) structures. Furthermore, each strut can comprise a
finer-scaled lattice, with a bending or stretching response at this
lower length scale~\cite{Fleck2010}. The following cases are
considered in this study:
\begin{itemize}
\item[(i) \textit{stretching lattice:}] A triangular lattice possesses
  a sufficiently high nodal connectivity ($Z=6$) that the lattice is
  stretching-dominated. Provided each strut is straight, deformation
  of the lattice induces stretching of each strut and thus we shall
  refer to the topology as stretching on the lattice scale as well as
  stretching on the strut scale. Such a lattice has a high modulus and
  inherits the ductility of the cell wall material~\cite{Gu2018}.
\item[(ii) \textit{stretching-bending lattice:}] Now consider a
  stretching-dominated lattice such as the triangular lattice, with
  struts which have enhanced axial compliance due to waviness of the
  struts. We refer to such a lattice as a stretching-bending lattice.
\item[(iii) \textit{bending lattice:}] A hexagonal lattice with
  straight struts deforms by bending of the struts. The presence of
  strut waviness has a negligible effect upon the bending stiffness of
  the strut and thereby has a negligible effect upon the macroscopic
  compliance. We refer to this lattice as a bending lattice.
\end{itemize}

\subsection{Imperfections in struts}
\label{sec:imperfections}
The sensitivity of modulus, strength, and ductility to imperfections
within a foam or a lattice has been studied systematically, see for
example \cite{Gibson1999, Ronan2016, Symons2008, Schmidt2001,
  Tankasala2015, Tankasala2017, Fleck2007, Romijn2007, Grenestedt2005,
  Liu2017,Tankasala2020}. Imperfections include missing struts,
misaligned struts, misplaced joints, Plateau borders and
cell-level inclusions for the case of metallic
foams~\cite{Ronan2016}. Recently, the sensitivity of the dispersion in
macroscopic properties to the statistical distribution of
imperfections has been analysed for brittle~\cite{Seiler2019} and
visco-plastic honeycombs~\cite{Seiler2019a} made by rapid prototyping:
a scatter in strut thickness and in strut ductility have a major
detrimental effect on the macroscopic strength.

\subsection{Influence of strut waviness on macroscopic properties}
Wavy struts can have a profound influence on the macroscopic
properties of lattice materials: lattices comprising wavy struts allow
tuneable Poisson's ratio~\cite{Chen2017} and macroscopic
stiffness~\cite{Jang2015}. Symons et al.~\cite{Symons2008} and
Grenestedt~\cite{Grenestedt1998} predicted the influence of
sinusoidal strut waviness on the macroscopic stiffness of triangular,
stretching-dominated lattices. The axial stiffness $k$ of a strut with
such waviness of amplitude $a$ is~\cite{Symons2008}
\begin{equation}
  \label{eq:strut_stiffness_wavy}
  k = \frac{E_\mathrm{s}t}{\ell}\frac{1}{1+6(a/t)^2}
\end{equation}
where $t$ is the strut thickness, $\ell$ the node-to-node strut length
and $E_\mathrm{s}$ is the axial modulus of the parent material. Thus,
a waviness amplitude $a/t=2$ leads to a knock-down in the axial
stiffness of the strut by a factor of 25 and consequently there is a
similar knock-down in the overall modulus of a triangular lattice
comprising such wavy struts. While waviness reduces lattice stiffness
it can enhance lattice ductility. For example, Ma et
al.~\cite{Ma2016,Ma2016a} and Jang et al.~\cite{Jang2015} investigated
polyimide lattices materials comprising horseshoe-shaped struts
embedded in a soft polymeric matrix. The inherent waviness of the
horseshoe shaped struts significantly enhanced the ductility of these
lattices.

\subsection{Scope of study}
The aim of the current study is to investigate lattice designs that
deliver desired ductilities and ultimate strengths. We constructed
two-dimensional (2D) steel lattice materials of constant relative
density $\bar\rho=0.1$ and made from wavy struts. Detailed measurements of the
tensile responses of the lattice materials, their manufacturing
induced defects and their constituent materials are reported. The
measurements are accompanied by FE simulations that include the
observed manufacturing defects. The combined effect of strut waviness
and manufacturing defects is mapped out by FE simulations for both the
bending-dominated and stretching-dominated lattices to give guidelines
for the design of lattice materials comprising wavy struts.

\section{A preliminary assessment of the significance of undercut and
  waviness}
\label{sec:prel-assessm-sign}
\begin{figure}[htbp]
  \centering
  \setlength{\labelsep}{0.1cm}
  \sidesubfloat[]{\includegraphics[width=0.88\textwidth]{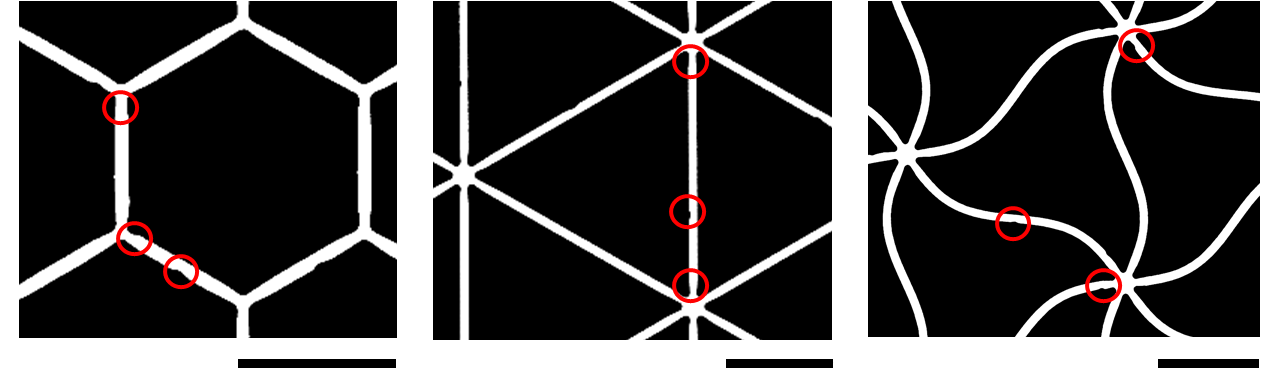}\label{fig:fs2a}}\\[\baselineskip]
  \sidesubfloat[]{\includegraphics[width=0.88\textwidth]{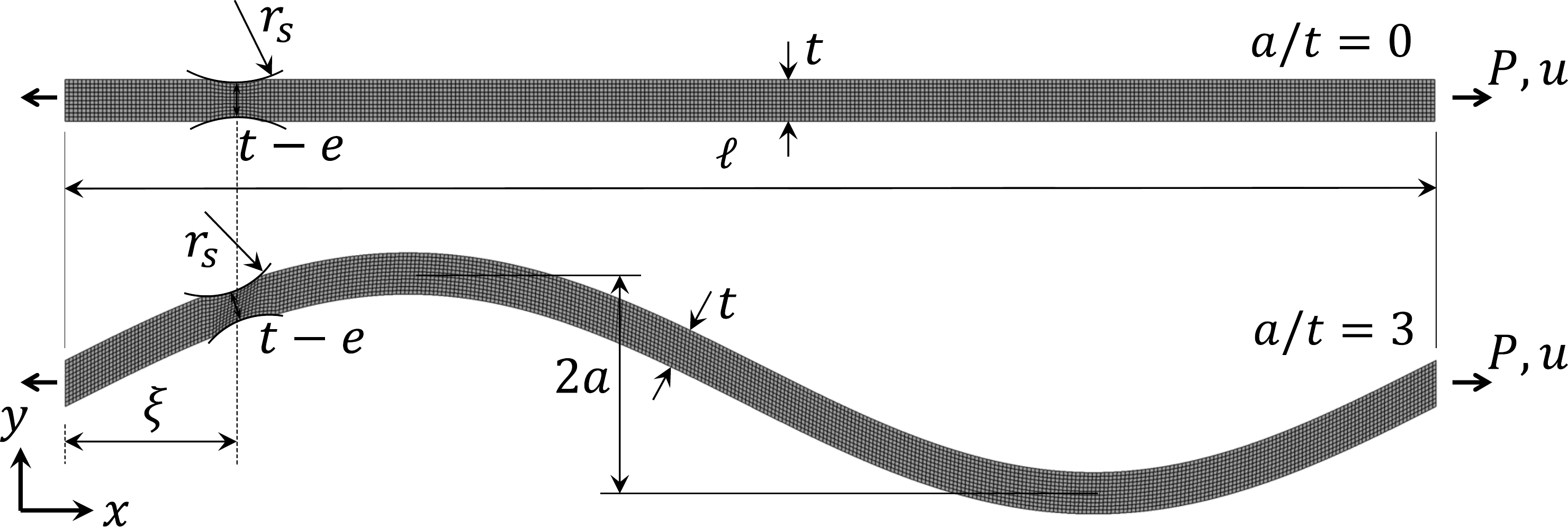}\label{fig:fs2b}}\\[\baselineskip]
  \setlength{\labelsep}{-0.4cm}
  \sidesubfloat[]{\includegraphics[width=0.42\textwidth]{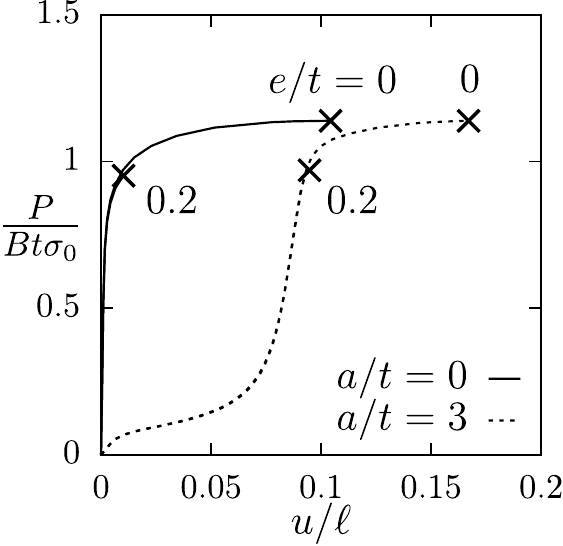}\label{fig:fs2c}}\hspace{0.5cm}
  \sidesubfloat[]{\includegraphics[width=0.43\textwidth]{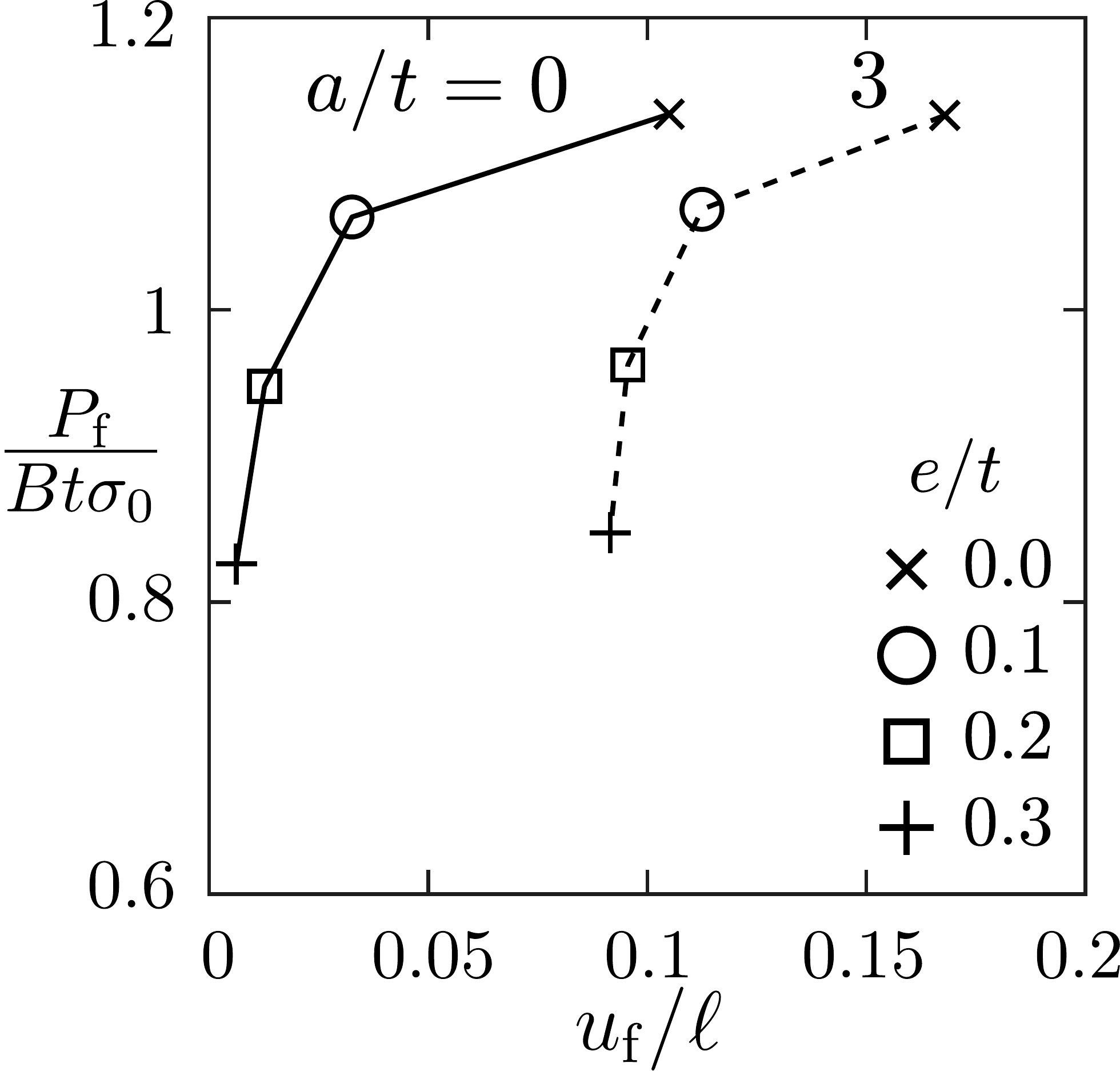}\label{fig:fs2d}}
  \caption{\subref{fig:fs2a} Computerised tomography (CT) scan images
    showing the observed undercuts (marked with red circles) in
    various samples (the scale bar is of length
    \SI{10}{mm}). \subref{fig:fs2b} FE model for straight and wavy
    struts with undercuts. The various geometric parameters are
    labelled and applied loading shown.  \subref{fig:fs2c} FE
    predictions of the force versus displacement response for straight
    ($a/t=0$) and wavy ($a/t=3$) struts with undercuts of depth
    $e/t=0.0$ and $0.2$ for struts of aspect ratio $t/\ell=0.3$.
    \subref{fig:fs2d} FE predictions of the failure strength
    $P_\mathrm{f}$ versus failure displacement $u_\mathrm{f}$ for the
    struts in \subref{fig:fs2b} for 2 choices of $a/t$ and selected
    values of normalised undercut depth $e/t$. The undercut geometric
    parameters $r_\mathrm{s}/t=0.5$ and $\xi/\ell=1/8$ were used in
    all calculations.}
 \label{fig:undercut}
\end{figure}

A major focus of this study is to understand the interplay between
as-designed ``imperfections'' such as strut waviness, and typical
as-manufactured defects. A preliminary examination of steel
hexagonal and triangular lattices manufactured by water-jet cutting
revealed the presence of undercuts near joints; see the 3-dimensional
(3D) computerised tomography (CT) scan images shown in~\cref{fig:fs2a}
for both triangular and hexagonal lattices. Here we report a finite
element (FE) assessment of these undercuts by
modelling their effect on the tensile response of a single straight
strut or wavy strut.

Consider a two-dimensional (2D) strut of length $\ell$, in-plane
thickness $t$, sinusoidal waviness of amplitude $a$ and out-of-plane
thickness $B$ (\cref{fig:fs2b}). We introduce an undercut into this
strut at a distance $\xi$ from one end of the strut, with the undercut
characterised by its radius $r_\mathrm{s}$ and depth $e$ as shown in
\cref{fig:fs2b}. The plane strain tensile response of the strut was
analysed via FE simulations using the commercial finite element
package \textsc{Abaqus}. The strut was discretised by quadratic
elements (CPE8 in the \textsc{Abaqus} notation) with 10 elements
across the strut thickness. The solid material is idealised by a
deformation theory solid with a tensile stress versus strain response
parameterised by the Ramberg-Osgood relationship
\begin{equation}
\frac{\varepsilon}{\varepsilon_0}= \frac{\sigma}{\sigma_0} + \alpha \left(\frac{\sigma}{\sigma_0}\right)^{1/N}
\end{equation}
with the choice of parameters $\varepsilon_0=0.002$, reference strength
$\sigma_0=\SI{400}{MPa}$, $\alpha=5$ and hardening exponent $N=0.1$. A
monotonically increasing axial displacement $u$ was applied to the
strut until the conjugate load $P$ reached a peak value associated
with necking of the strut. Peak load defines the onset of failure
of the strut.

Predictions of the normalised load $P/(Bt\sigma_0)$ versus normalised
displacement $u/\ell$ are included in \cref{fig:fs2c} for a strut of
aspect ratio $t/\ell=0.03$. The figure includes predictions for both a
straight strut ($a/t=0$) and a wavy strut of sinusoidal shape with a
wavelength $\ell$ and amplitude $a$, such that $a/t=3$. Results are
given for two choices of normalised undercut
depth $e/t=0$ and $e/t=0.2$, with the normalised undercut radius and
position held fixed at $r_\mathrm{s}/t=0.5$ and $\xi/\ell=1/8$,
respectively. 

First, consider the case with no undercut ($e/t=0$). The
straight strut has an initial sharply rising load versus displacement
response. The response then displays a plateau as the strut material
undergoes plastic deformation. In contrast, the wavy strut has a
sigmoidal load-displacement response: initially, the wavy strut
straightens by bending. Thereafter, the response is
similar to that of the straight strut. The predictions in both cases are
terminated at peak load (marked by a cross) corresponding to the
onset of necking of the strut.

Second, consider the case of the struts with an undercut $e/t=0.2$. The
load-displacement response up to the onset of necking is identical to
that with no undercut. The undercut induces early necking at the
location of the undercut and the location of the peak load is marked
by the cross on the curves in \cref{fig:fs2c}. We define the peak load
$P_\mathrm{f}$ as the failure load and the corresponding displacement
$u_\mathrm{f}$ as the failure displacement. A cross-plot of the
normalised failure load $P_\mathrm{f}/(Bt\sigma_0)$ versus the
normalised failure displacement $u_\mathrm{f}/\ell$ is included in
\cref{fig:fs2d} for both the straight strut and strut with $a/t=3$,
for selected undercut depths $e/t$; $r_\mathrm{s}/t$ and
$\xi/\ell$ are held fixed at 0.5 and $1/8$, respectively. The failure load
is largely unaffected by waviness and increases with decreasing
undercut depth. However, the failure displacement for a given undercut
depth increases sharply with increasing waviness and also increases
with decreasing undercut depth (corresponding to the increase in the
failure load). These results are insensitive to the choice of undercut
radii and locations over a broad range of values
($r_\mathrm{s}/t=0.5-2.0$ and $\xi/\ell=0.1-0.9$). We proceed to use
this basic understanding to investigate first experimentally and then
numerically the design of lattice materials with prescribed strut
waviness.

\section{Experimental programme}
\label{sec:experimental_programme}
\begin{figure}[htbp]
  \centering
  \sidesubfloat[]{\includegraphics[width=0.75\textwidth]{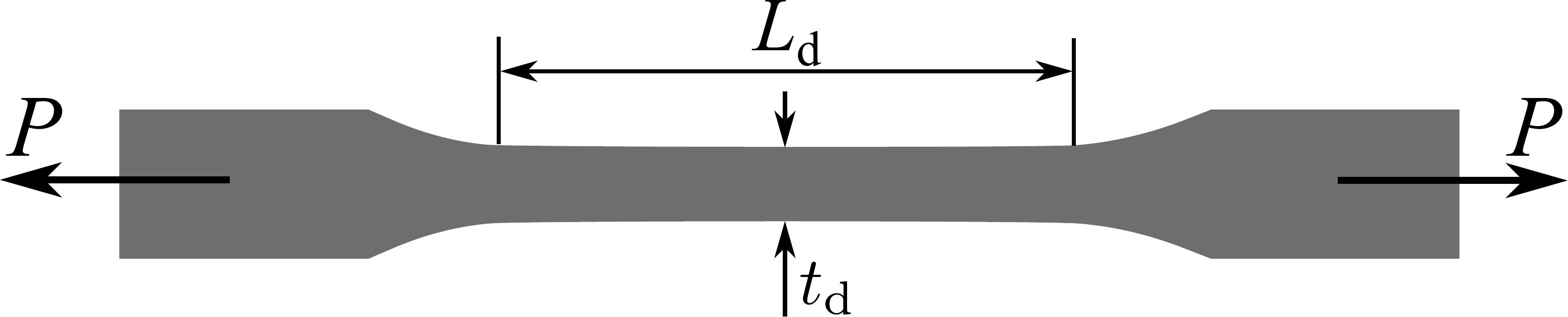}\label{fig:dbtension}}\\[\baselineskip]
  \setlength{\labelsep}{-0.35cm}
  \sidesubfloat[]{\includegraphics[width=0.4\textwidth]{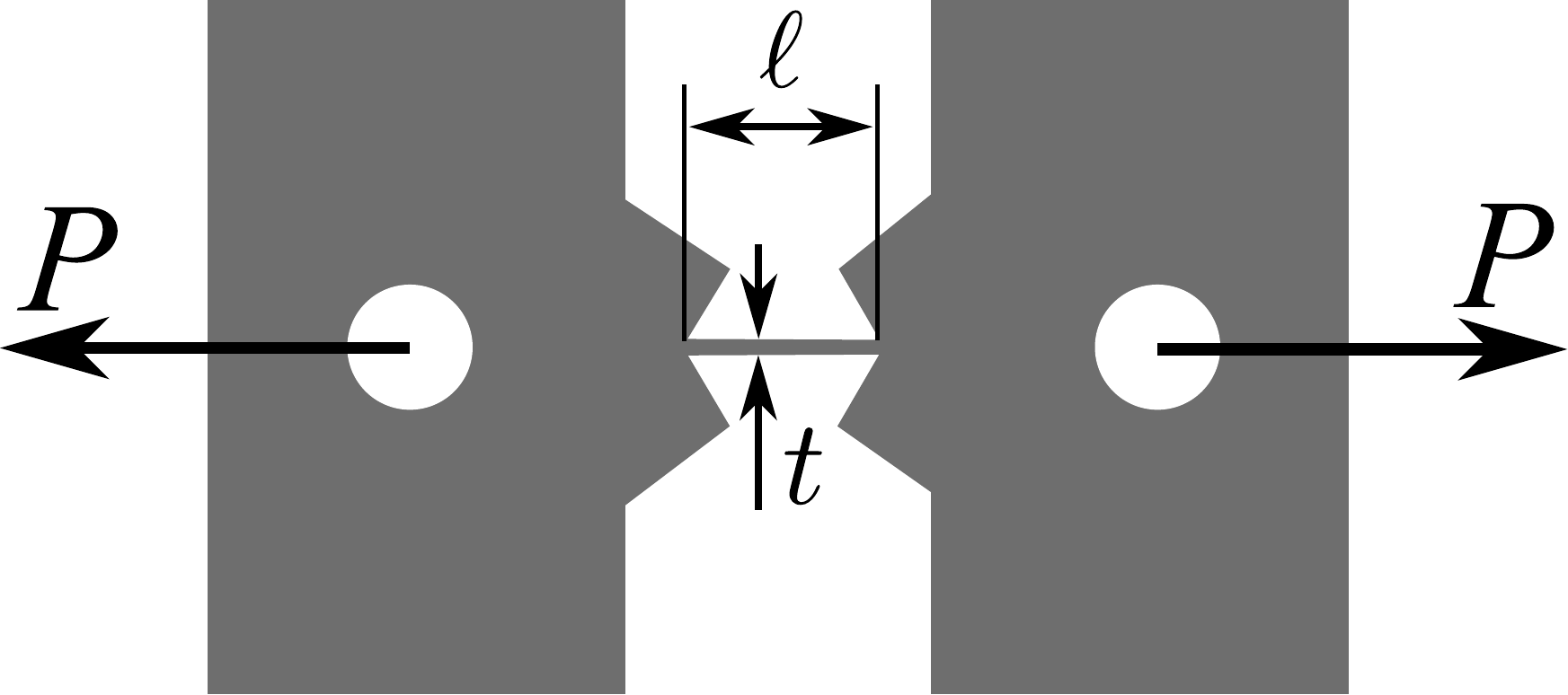}\label{fig:singleStrutTensionTria}}\hspace{1.0cm}
  \sidesubfloat[]{\includegraphics[width=0.4\textwidth]{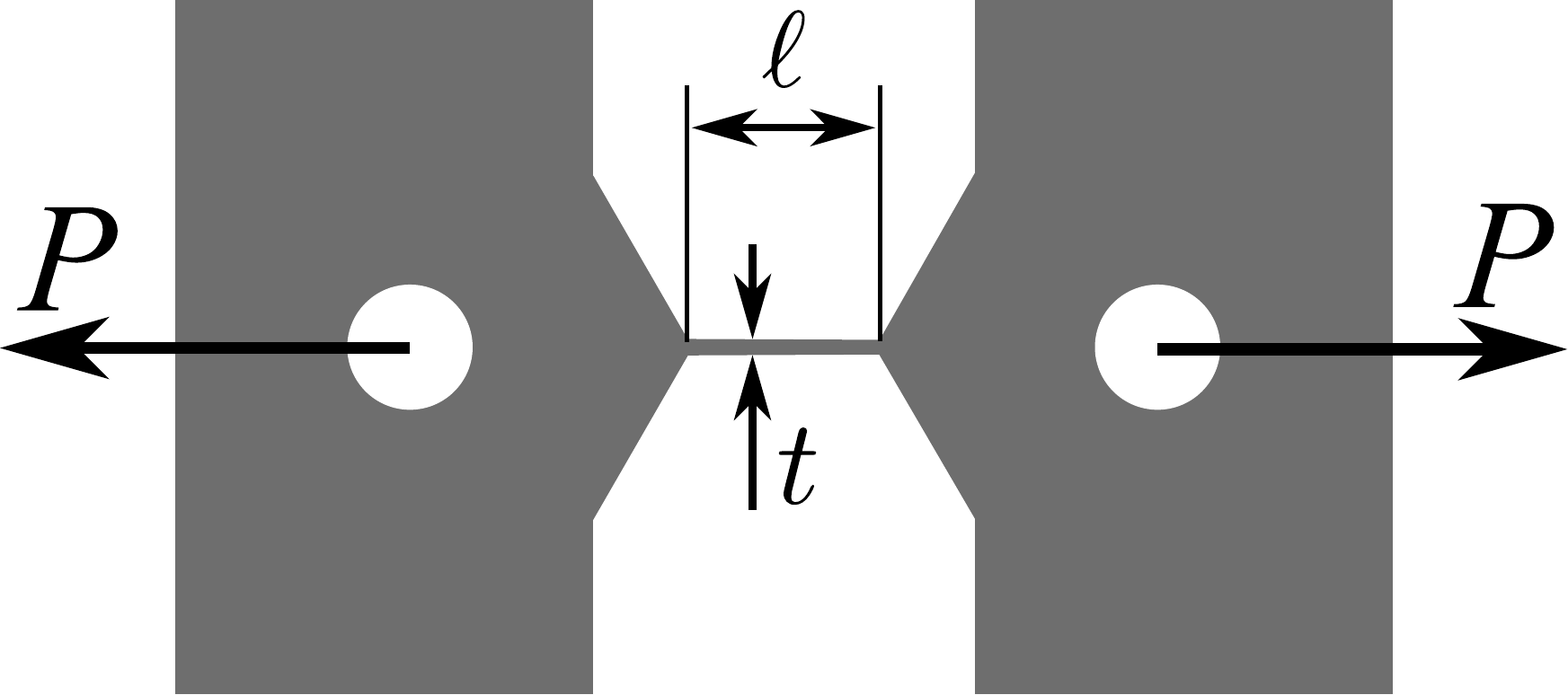}\label{fig:singleStrutTensionHex}}\\[\baselineskip]
  \setlength{\labelsep}{-0.25cm}
  \sidesubfloat[]{\includegraphics[width=0.5\textwidth]{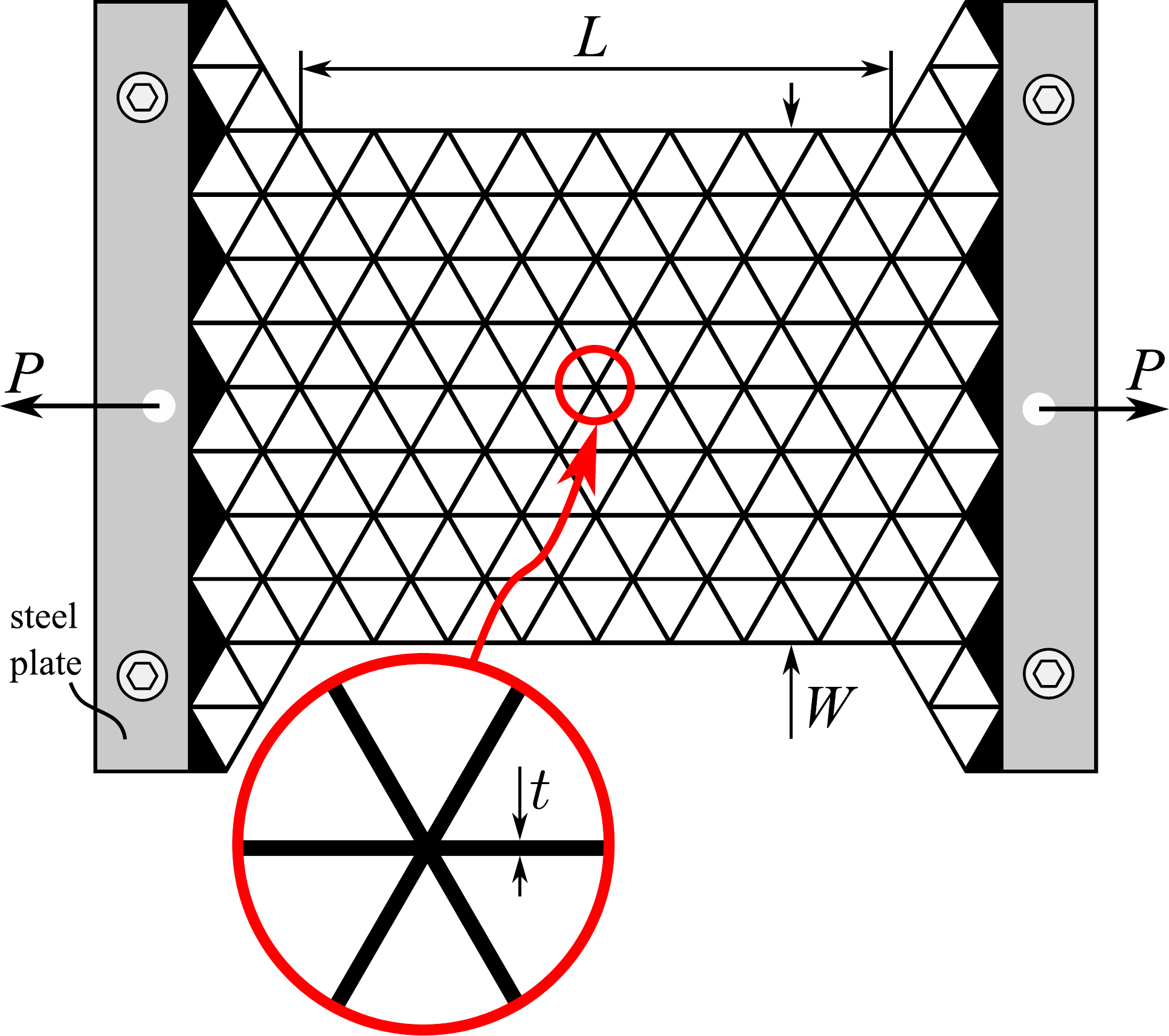}\label{fig:sketch_tria}}\\[\baselineskip]
  \sidesubfloat[]{\includegraphics[width=0.54\textwidth]{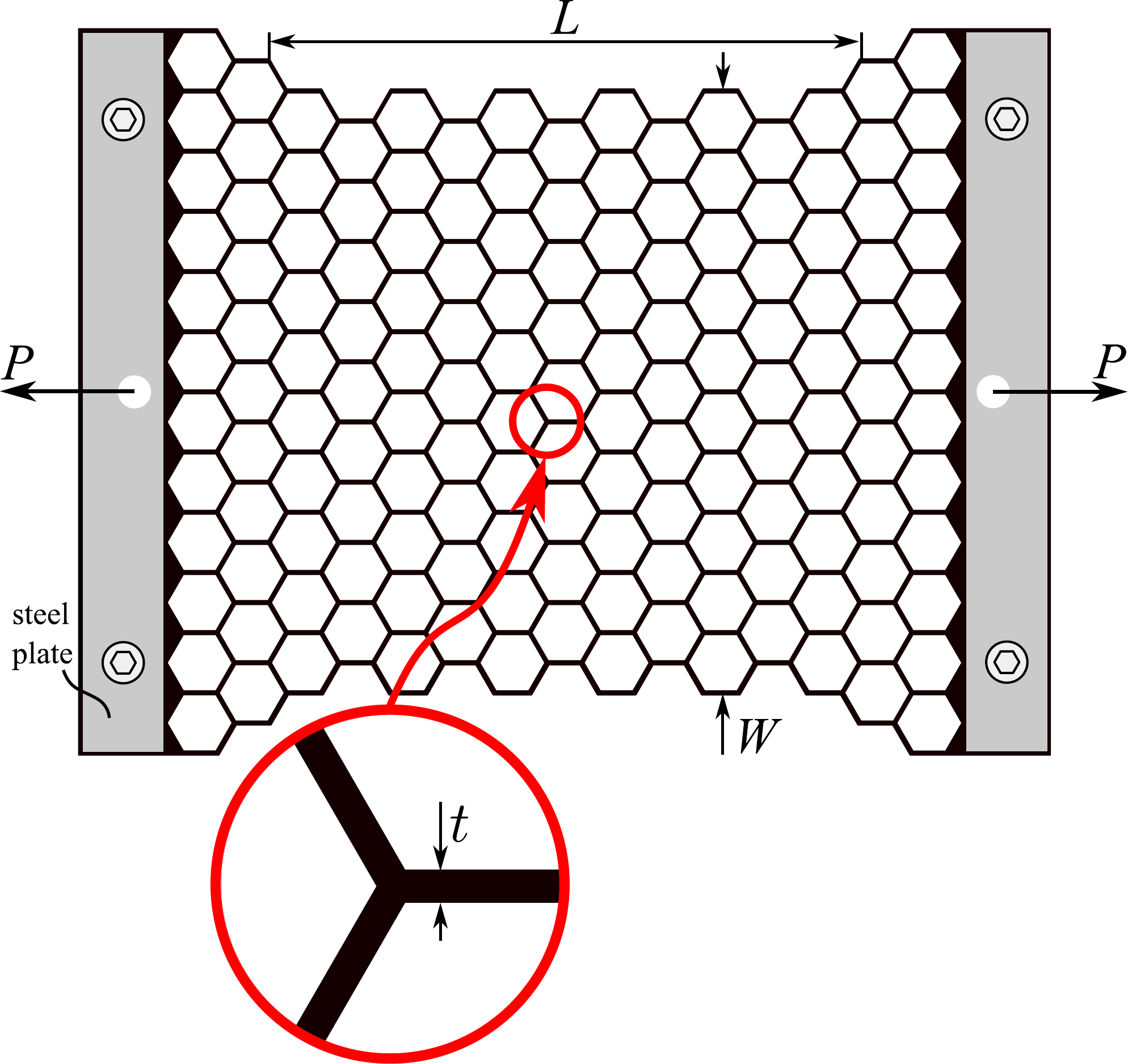}\label{fig:sketch_hex}}
  \caption{Geometry of specimens tested in this study.
    \subref{fig:dbtension} Dogbone specimen for tensile properties of
    present material. Single strut specimens mimicking struts in
    \subref{fig:singleStrutTensionTria} triangular and
    \subref{fig:singleStrutTensionHex} hexagonal lattices. The
    \subref{fig:sketch_tria} triangular and \subref{fig:sketch_hex} hexagonal
    lattice specimens. Leading dimensions are labelled on each of the
    sketches.}
  \label{fig:geometries_samples}
\end{figure}

Specimens were manufactured by water-jet cutting of hot-rolled
$B=3$\,mm thick steel sheets of grade \textit{S275} (low carbon steel
with a maximum of 0.25\% of carbon by weight) and hardness
185HV30. Three different specimen types were employed: (i) macroscale
dogbone specimens (\cref{fig:dbtension}) of the parent material to
characterise the solid material properties; (ii) specimens that
replicate the geometry of single struts within the lattices
(\cref{fig:singleStrutTensionTria,fig:singleStrutTensionHex}) and
(iii) triangular and hexagonal lattice specimens
(\cref{fig:sketch_tria,fig:sketch_hex}) of relative density
$\bar{\rho}=0.1$.

The tensile response at a nominal tensile strain rate of
$2\times 10^{-4}$~s$^{-1}$ was measured using a screw-driven test
machine, with the load cell of the machine used to measure the applied
tensile load $P$. A laser extensometer was used to measure the gauge
section extension of the dogbone parent material specimens while
Digital Image Correlation (DIC) was used to measure displacements in
the single strut and lattice specimens.

\subsection{Geometry of struts and the single strut specimens}
\label{sec:strut_geometries}
\begin{figure}[htbp]
  \centering
  \includegraphics[width=0.44\textwidth]{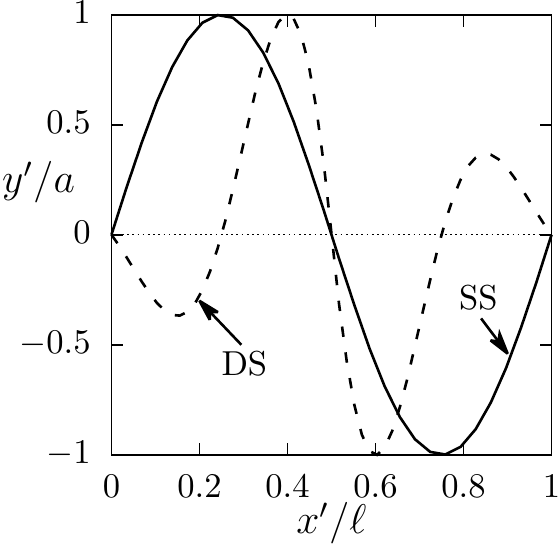}
  \caption{Shape of the sinusoidal (SS) and decaying sinusoidal (DS)
    struts as parameterised by \cref{eq:sine_shape} and
    \cref{eq:double_sine_shape}, respectively.}
 \label{fig:sineFunc}
\end{figure}

The struts of the lattices investigated here were either straight (S)
or wavy. Two wavy geometries were employed: (i) a single sinusoidal
(SS) waveform and (ii) a \ds sinusoidal (DS) waveform. These waveforms
are parameterised as follows. In the local Cartesian co-ordinate system
$(x',y')$, the equation parameterising the SS waveform is
\begin{equation}
  \label{eq:sine_shape}
  y'=a\sin\left(\frac{2 \pi x'}{\ell}\right)
\end{equation}
where $a$ is the amplitude of the wavy strut, and the wavelength
$\ell$ is the distance between the end points of the strut; see
\cref{fig:sineFunc}. The DS waveform retains the symmetry of the SS
waveform about the mid-span of the strut and is parameterised by
\begin{equation}
  \label{eq:double_sine_shape}
  y'=1.6\,a\,\sin\left(\frac{4 \pi x'}{\ell}\right) \exp\left(\frac{-4x'}{\ell}\right)
\end{equation}
where the factor of $1.6$ has been included to ensure that the maximum
amplitude of the waviness equals $a$ as seen in \cref{fig:sineFunc}.

In order to investigate the tensile responses of these struts within
the lattices, we also manufactured single strut specimens as shown in
\cref{fig:singleStrutTensionTria,fig:singleStrutTensionHex}. These
specimens comprise either straight or wavy struts and comprised the
same node geometries as are present in the triangular and hexagonal
lattices.

\subsection{Lattice specimens}
\label{sec:lattice}
The tensile response of triangular and hexagonal lattices comprising
straight and wavy struts was investigated using dogbone shaped
specimens (\cref{fig:sketch_tria,fig:sketch_hex}) to ensure that
failure occurred within the gauge section. The gauge section of the
triangular and hexagonal lattice specimens comprised $8\times 8$ and
$11\times 10$ cells, respectively. The specimens were bolted to
3\,mm thick steel end tabs to enable gripping of the specimens for
tensile loading. The specimens were manufactured by water-jet cutting
of the 3\,mm thick steel sheets, with the tensile loading direction of
the specimens aligned with the rolling direction of the steel
sheets. The radius of the water-jet nozzle was 0.34\,mm and thus the
corner radii of the nodes exceeds 0.34\,mm.

All lattice specimens investigated here had a relative density
$\bar\rho=0.1$. While the relative density of lattices with straight
struts is only a function of $t/\ell$, the magnitude of $\bar\rho$ for
lattices with wavy struts depends strongly on the strut shape. It is
instructive to define the arc-length $\ell_\mathrm{s}$ of a strut with
node-to-node length $\ell$ as
\begin{equation}
  \label{eq:arc_length_sine}
  \ell_\mathrm{s}=\int^{\ell}_0 \sqrt{1+\left( \frac{dy'}{dx'} \right)^2}\, dx'
\end{equation}
where $(x',y')$ is the local, Cartesian co-ordinate system and the
$x'$-direction is co-incident with a straight strut between the
nodes. The modified form of \cref{eq:relative_dens_straight} for
lattices with wavy struts is then
\begin{equation}
  \label{eq:relative_dens_wavy}
  \bar{\rho}=A \frac{t\ell_\mathrm{s}}{\ell^2}
\end{equation}
For all lattices investigated here we kept $t=\SI{0.81}{mm}$, and
$t\ell_\mathrm{s}/\ell^2=0.03$ and 0.09 for the triangular and
hexagonal lattices, respectively, independent of the strut shape, so
that $\bar\rho=0.1$ in all cases. The specific geometric parameters of
all lattice geometries investigated here are listed in
\cref{tab:geometry_constRelD}.

While the water-jet cutting of the lattice used a computed aided
drawing (CAD) input\footnote{Software to generate the wavy lattice
  geometries: \cite{Seiler2020a}.} of the detailed specimen geometry
absent any defects, changes in the cutting speed of the water-jet as
it went around the corners of the lattice and residual stresses within
the steel sheet meant that the as-cut lattice did not precisely match
the CAD specification. X-ray CT examination of the
manufactured lattices revealed a dispersion in the strut thicknesses
$t$, a finite radius $r_\mathrm{n}$ of corners at nodes, and undercuts
within the struts near the nodes (see \cref{fig:fs2a}). The X-ray CT
images were used to characterise these defects by making measurements
over 280 struts in 14 different specimens and the findings of these
measurements are summarised as follows:

\begin{enumerate}
\item While the mean strut thickness at mid-span attained the
  specified value of 0.81\,mm, the strut thicknesses in each specimen
  were normally distributed with a standard deviation of
  \SI{0.04}{mm}.
\item The corner radii were also normally distributed, with a mean
  value $<r_\mathrm{n}>=0.5$\,mm and standard deviation of
  \SI{0.08}{mm}.
\item Nearly all struts had undercuts near the nodes, as characterised
  by a mean radius $<r_\mathrm{s}>\approx <r_\mathrm{n}>$ and undercut
  depths in the range $0.2\le e/t\le 0.3$; see \cref{fig:fs2b} for the
  definitions of $r_\mathrm{s}$ and $e$.
\end{enumerate}

\section{Material characterisation}
\label{sec:material_characterisation}

\subsection{Solid material response}
\label{sec:solid_mat_resp}
\begin{figure}[htbp]
  \includegraphics[width=0.50\textwidth]{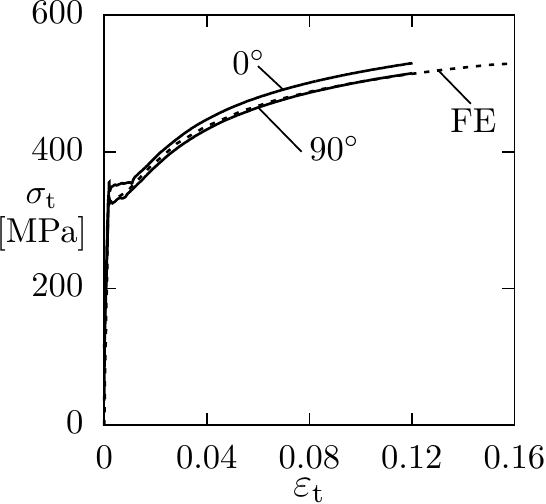}
  \caption{True stress versus strain response of the parent metal
    steel used to manufacture the specimens. The measured responses in
    the rolling direction of the steel sheet ($0^\circ$) and
    perpendicular to the rolling direction ($90^\circ$) as shown until
    the onset of necking. The response used in the FE calculations is
    also included as a dashed line.}
  \label{fig:stressStrainS275DB_oneStrainRate}
\end{figure}

The material properties of solid low-carbon steel sheets were measured
from the response of a large dogbone-shaped specimen of gauge
dimensions $L_\mathrm{d}=\SI{80}{mm}$ and $t_\mathrm{d}=\SI{10}{mm}$
(see \cref{fig:dbtension}) at \SI{0}{\degree} and \SI{90}{\degree} to
the hot-rolling direction of the steel sheets. The measured true
stress $\sigma_\mathrm{t}$ versus true strain $\varepsilon_\mathrm{t}$
responses of the solid dogbone specimens are shown in
\cref{fig:stressStrainS275DB_oneStrainRate}. All specimens respond in
a ductile manner with a negligible effect of the hot-rolling direction
upon the tensile response, such that the Young's modulus is
$E_\mathrm{s}=\SI{210}{GPa}$, yield strength is
$\sigma_\mathrm{YS}=338\pm\SI{12}{MPa}$, ultimate tensile strength is
$\sigma_\mathrm{UTS}=465\pm\SI{6}{MPa}$ and the nominal tensile
failure strain is $\varepsilon_\mathrm{fs}=0.24\pm0.003$. Over a
strain range $0.03<\varepsilon_\mathrm{t}<0.12$, the true stress
versus true strain response is well-approximated by
$\sigma_\mathrm{t}/\sigma_0 =
(\varepsilon_\mathrm{t}/\varepsilon_0)^N$ where $N=0.1$.

\subsection{Single strut response}
\label{sec:single_strut_resp}
\begin{figure}[htbp]
  \centering
  \setlength{\labelsep}{-0.30cm}
  \sidesubfloat[]{\includegraphics[width=0.47\textwidth]{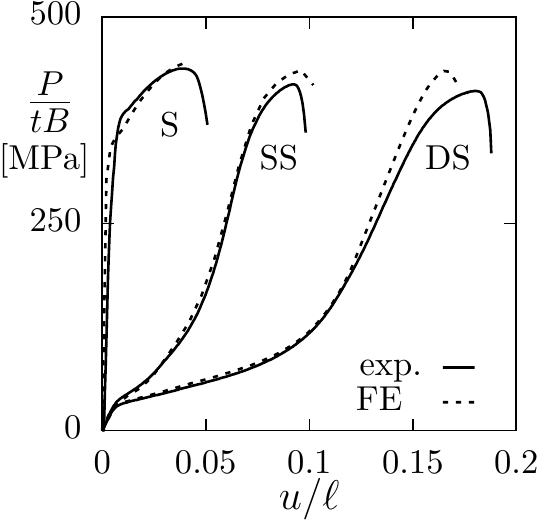}\label{fig:stressStrain_oldMat_tria_wavy}}\hfill
  \sidesubfloat[]{\includegraphics[width=0.47\textwidth]{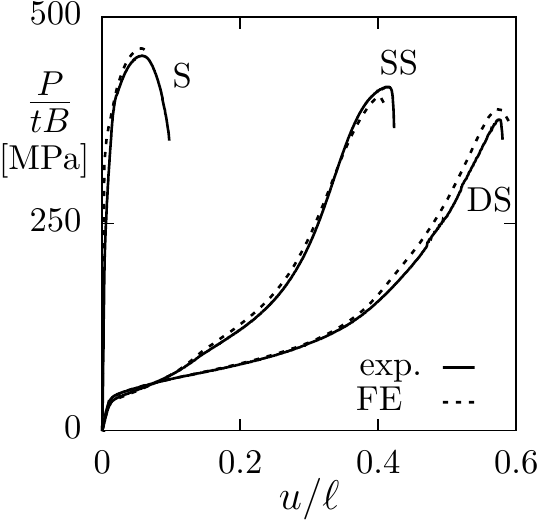}\label{fig:stressStrain_oldMat_hex_wavy}}
  \caption{The measured and predicted nominal stress versus strain
    responses of the single strut specimens mimicking struts in the
    \subref{fig:stressStrain_oldMat_tria_wavy} triangular and
    \subref{fig:stressStrain_oldMat_hex_wavy} hexagonal lattices. The
    wavy struts (SS and DS) have a normalised amplitude $a/t=2.7$ and
    the FE calculation employed an undercut of depth $e/t=0.1$.}
  \label{fig:stressStrain_singleStrut}
\end{figure}

The tensile responses of single struts, of the same geometry as that
used in the triangular and hexagonal lattices, are given in
\cref{fig:stressStrain_singleStrut}.  The measured nominal stress,
$P/(tB)$, defined in terms of the measured tensile load $P$, is
plotted in \cref{fig:stressStrain_singleStrut} as a function of
nominal strain $u/\ell$, where $u$ is the applied axial displacement
responses of the struts. In \cref{fig:stressStrain_oldMat_tria_wavy}
the responses are plotted for the $t\ell_\mathrm{s}/\ell^2=0.03$
struts that are representative of those in the triangular lattices
while in \cref{fig:stressStrain_oldMat_hex_wavy} we include
measurements for the $t\ell_\mathrm{s}/\ell^2=0.09$ struts that mimic
those in the hexagonal lattices. First, consider the straight struts
(S). It is evident that the tensile strength is comparable to the
solid material response. However, the ductility $u_\mathrm{f}/\ell$ of
straight struts is knocked down to 0.05 and 0.1 for the struts in the
triangular and hexagonal lattices, respectively, compared to the
parent material value of $\varepsilon_\mathrm{f} = 0.24$.  These
reduction in ductility of the straight single struts compared to that
of the solid material is due premature necking at the undercuts
introduced by water-jet cutting near the joints.

Strut waviness brings about a qualitative change to the response. Wavy
struts are first straightened before they neck and therefore waviness
increases the ductility of single struts, with the largest increase of
ductility exhibited by \ds sinusoidal struts (DS). However, the
ultimate tensile strength is largely unaffected by the presence of
waviness.

\section{ Finite Element calculations}
\label{sec:numerical_study}
\begin{figure}[htbp]
  \centering \setlength{\labelsep}{0.4cm}
  \sidesubfloat[]{\includegraphics[width=0.66\textwidth]{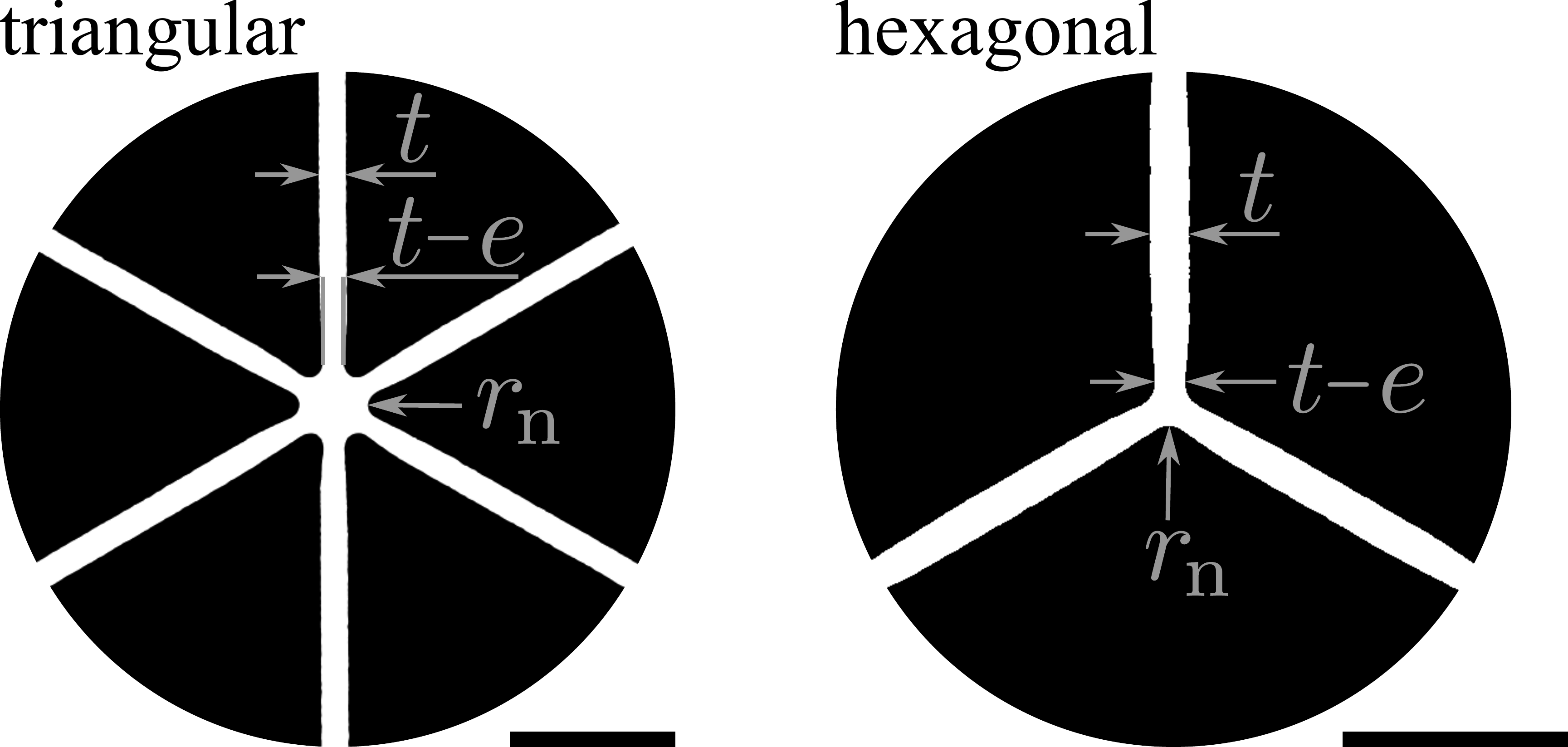}\label{fig:CT_nodes}}\\[\baselineskip]
  \setlength{\labelsep}{-0.3cm}
  \sidesubfloat[]{\includegraphics[width=0.71\textwidth]{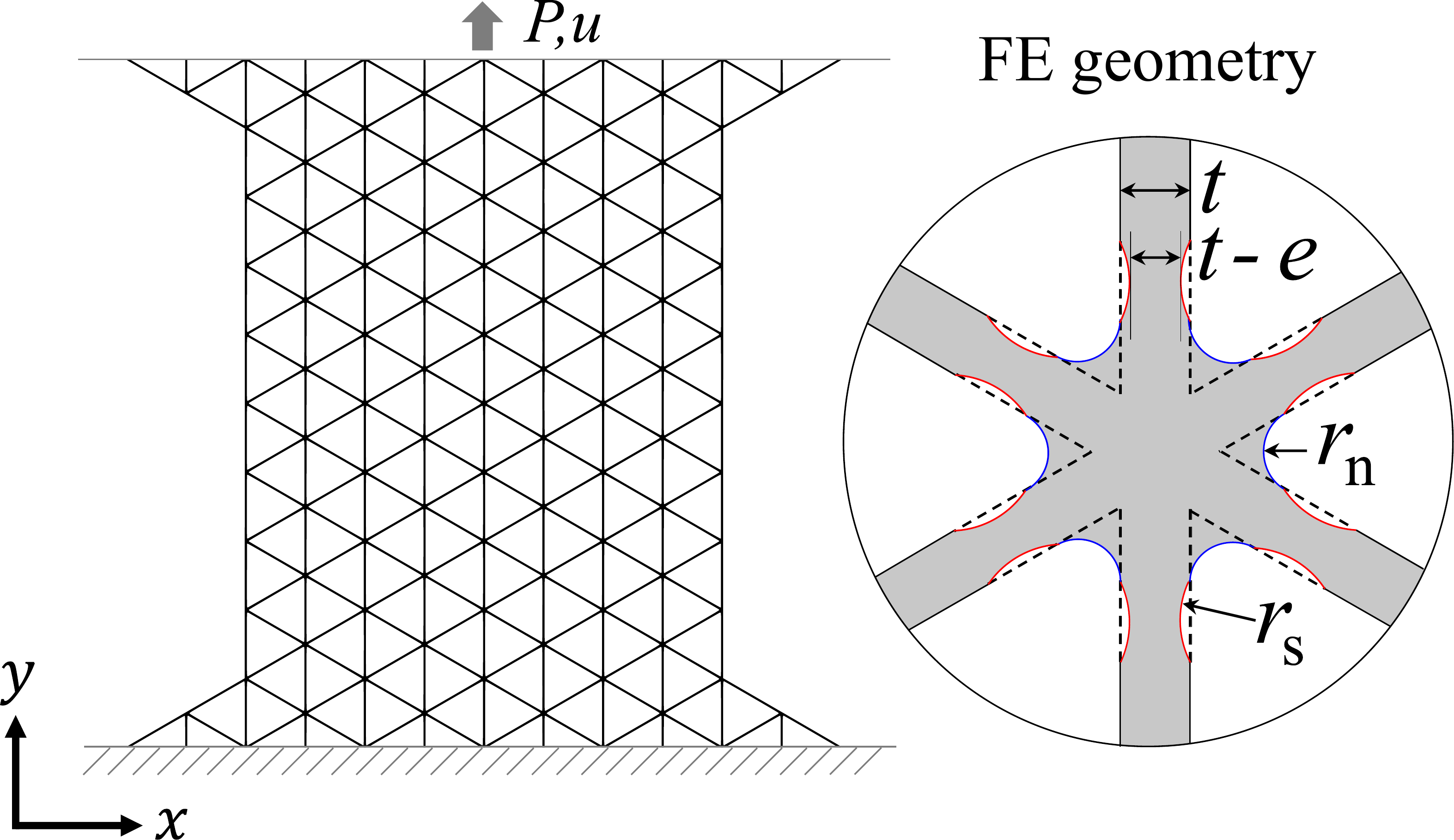}\label{fig:FEtri}}\\[\baselineskip]
  \sidesubfloat[]{\includegraphics[width=0.7\textwidth]{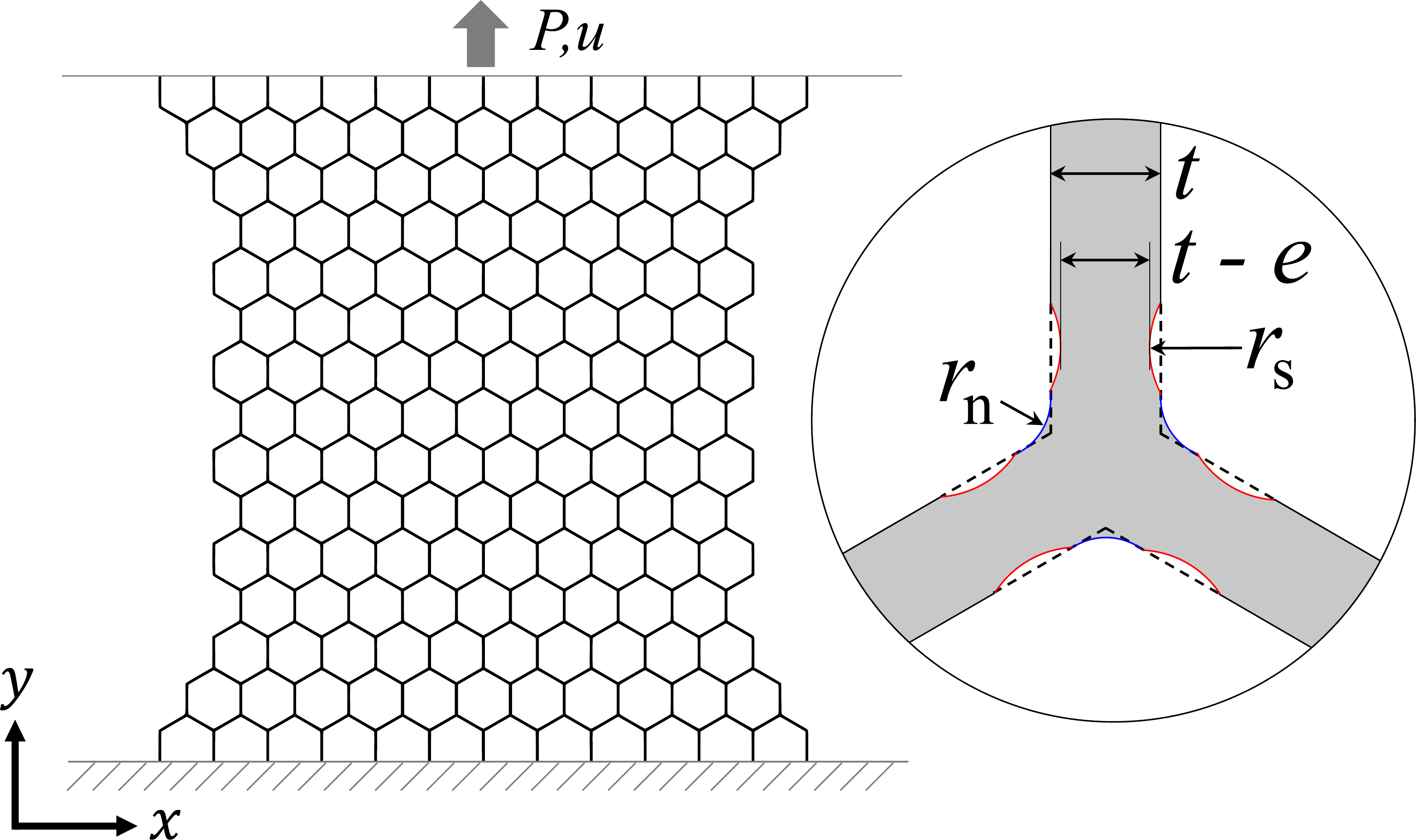}\label{fig:FEhex}}
  \caption{\subref{fig:CT_nodes} CT image of a node
    in triangular and hexagonal lattices (scale bar is of size
    \SI{5}{mm}). Geometry, loading and boundary conditions employed in
    the FE simulations of \subref{fig:FEtri} triangular and
    \subref{fig:FEhex} hexagonal lattice specimens. The insets in \subref{fig:FEtri} and \subref{fig:FEhex}
    show details of a node to illustrate their geometry and the
    imperfections in the form of an undercut.}
 \label{fig:FEdetails}
\end{figure}

Static finite element (FE) simulations were performed using
\textsc{Abaqus}/Standard v2018 to simulate the tensile response of the
single struts and the uniaxial tensile response of the triangular and
hexagonal lattices. The 2D plane strain geometry in the FE models
mimicked the as-manufactured specimens as observed in the CT
images. All struts were ascribed a thickness equal to the mean
measured value, $t=0.81$\,mm. In addition, each node of the lattice
was assumed to have a corner radius $r_\mathrm{n}=\SI{0.5}{mm}$ and an
undercut of radius $r_\mathrm{s}=\SI{0.5}{mm}$. The detailed node
geometries were consistent with the CT images, with representative
examples for the triangular and hexagonal lattices shown in
\cref{fig:CT_nodes}. The corresponding FE geometries of the lattice
specimens, with details of the nodes shown in insets, are included in
\cref{fig:FEtri,fig:FEhex} for the triangular and hexagonal lattices,
respectively. The undercut depth $e/t$ varied significantly between
specimens and the measured values were used in the simulations; their
values are explicitly specified in the presentation of the numerical
predictions.

The FE mesh of the lattice comprises rectangular elements with
quadratic shape functions (CPE8 in \textsc{Abaqus} notation). At least
4 elements across the thickness of each strut were present in order to
capture the stress concentration due to the nodes and the
undercuts. Uniaxial loading of the lattice specimens was simulated by
constraining all degrees of freedom along the bottom edge of the
specimen while the top edge is subjected to uniform displacement in
the $y$-direction of the specimen, with the $x$-direction displacement
of those nodes constrained to be zero, see \cref{fig:FEtri,fig:FEhex}.
The solid material was modelled as a J2 flow theory solid with Young's
modulus $E_\mathrm{s}=\SI{210}{GPa}$, Poisson's ratio $\nu=0.3$ and a
true tensile stress versus plastic strain response given by the
measurements in \cref{fig:stressStrainS275DB_oneStrainRate}. No damage
model was employed in the FE simulations with failure assumed to arise
from necking of the struts.

\subsection{FE predictions of the tensile response of a single strut}
We validated the FE model by comparing predicted and measured single
strut responses. The predictions employed a FE model with the single
strut modelled in an identical manner to the struts within the lattice
specimens. The FE predictions of the tensile responses of the single
strut are included in
\cref{fig:stressStrain_oldMat_tria_wavy,fig:stressStrain_oldMat_hex_wavy}
for an undercut depth $e/t=0.1$, as measured by X-ray CT.  Excellent
agreement is observed in all cases including the onset of softening
due to necking. This demonstrates the fidelity of the FE model and
validates the assumption of not including damage mechanisms in the
solid material.

\section{Tensile response of lattice specimens: predictions versus
  experiment}
\label{sec:tensile_response_lattices}
We proceed to present both measurements and FE predictions of the
tensile responses of the lattice specimens. Results are presented in
terms of a nominal stress $P/(W B)$ and nominal strain $u/L$ where the
specimen gauge width $W$ and gauge length $L$ are defined in
\cref{fig:sketch_tria,fig:sketch_hex}, while $P$ and $u$ are the
applied tensile load and corresponding extension of the gauge length
of the specimen, respectively.

\begin{figure}[htbp]
  \centering
  \setlength{\labelsep}{-1.cm}
  \sidesubfloat[]{\includegraphics[width=0.5\textwidth]{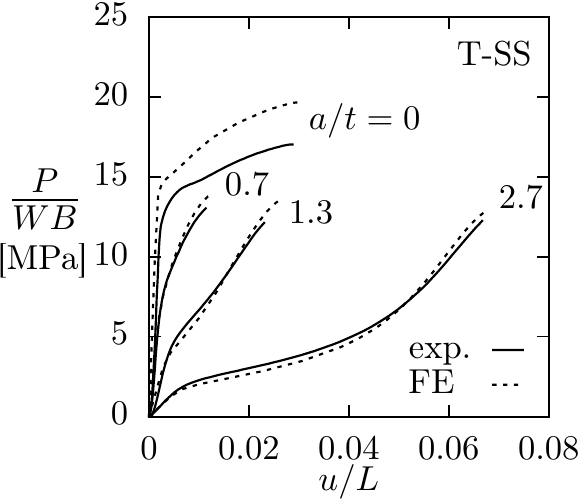}\label{fig:stressStrain_constRelD_tria_sine_at}}\hfill
  \sidesubfloat[]{\includegraphics[width=0.5\textwidth]{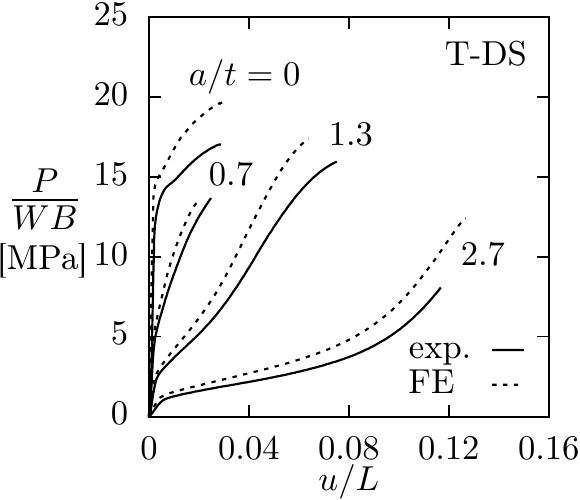}\label{fig:stressStrain_constRelD_tria_doublesine_at}}\\[\baselineskip]
  \sidesubfloat[]{\includegraphics[width=0.5\textwidth]{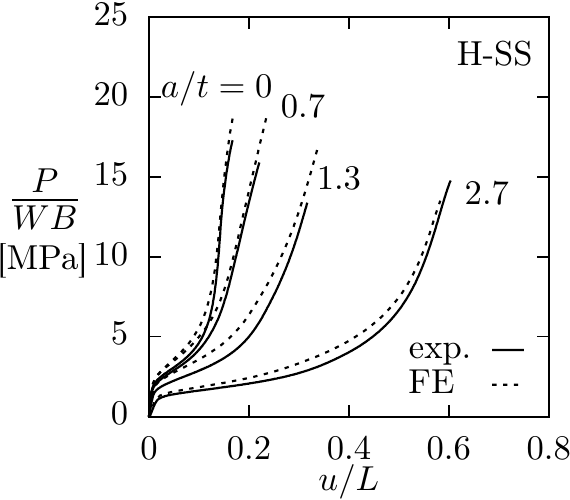}\label{fig:stressStrain_constRelD_hex_sine_at}}\hfill
  \sidesubfloat[]{\includegraphics[width=0.5\textwidth]{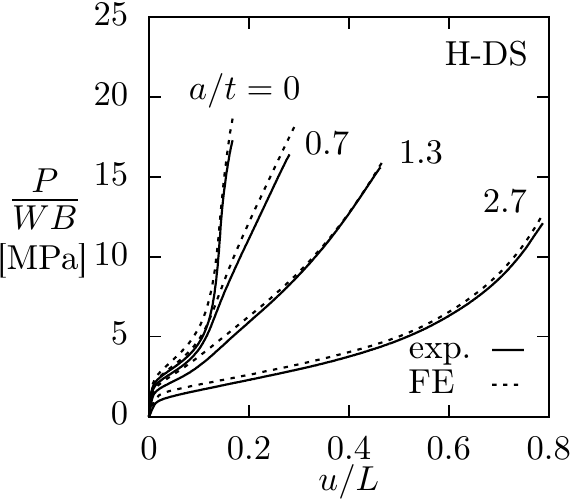}\label{fig:stressStrain_constRelD_hex_doublesine_at}}
  \caption{Measurements and predictions of the stress versus strain
    response of the lattices with sinusoidal (SS) and decaying
    sinusoidal (DS) shaped
    struts. \subref{fig:stressStrain_constRelD_tria_sine_at}
    Triangular lattices with sinusoidal strut shape (T-SS);
    \subref{fig:stressStrain_constRelD_tria_doublesine_at} triangular
    lattices with \ds sinusoidal strut shape (T-DS);
    \subref{fig:stressStrain_constRelD_hex_sine_at} hexagonal lattices
    with sinusoidal strut shape (H-SS), and
    \subref{fig:stressStrain_constRelD_hex_doublesine_at} hexagonal
    lattices with \ds sinusoidal strut geometry (H-DS). The FE
    calculations use $e/t$ values listed in
    \cref{tab:geometry_constRelD} for each specimen.}
  \label{fig:stressStrain_constRelD_combined2}
\end{figure}

\begin{figure}[htbp]
  \centering
      \setlength{\labelsep}{0.25cm}
    \sidesubfloat[]{\includegraphics[width=0.99\textwidth]{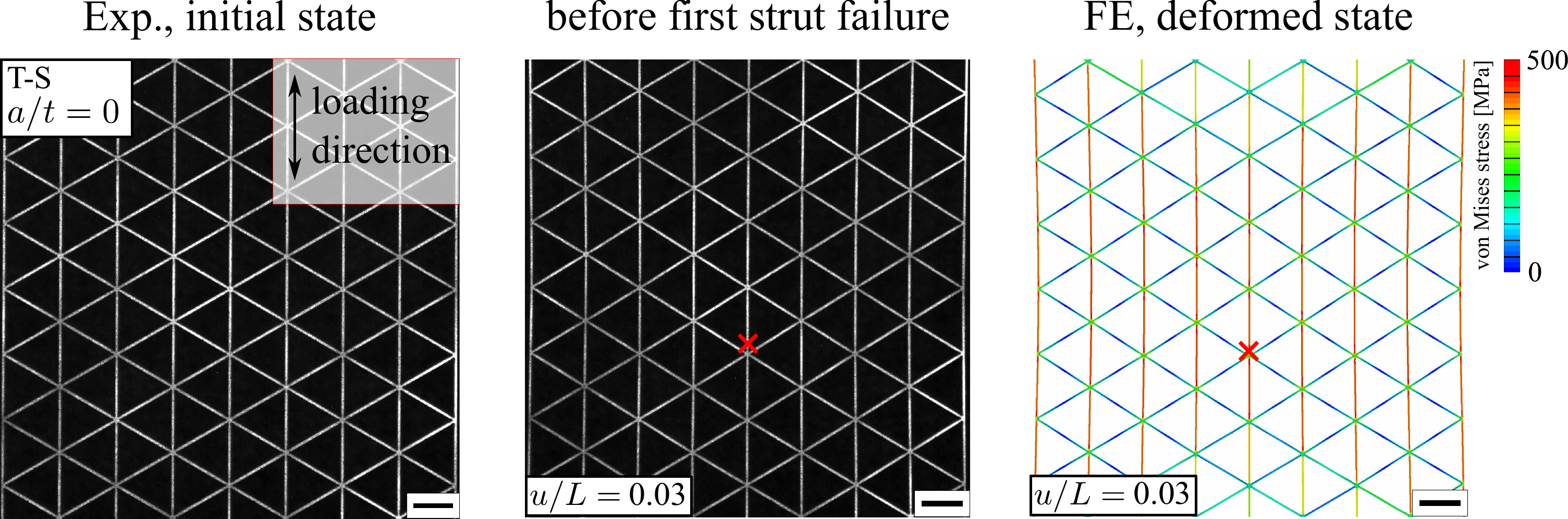}\label{fig:failureSequence_tria_a}}\\[\baselineskip]
    \sidesubfloat[]{\includegraphics[width=0.99\textwidth]{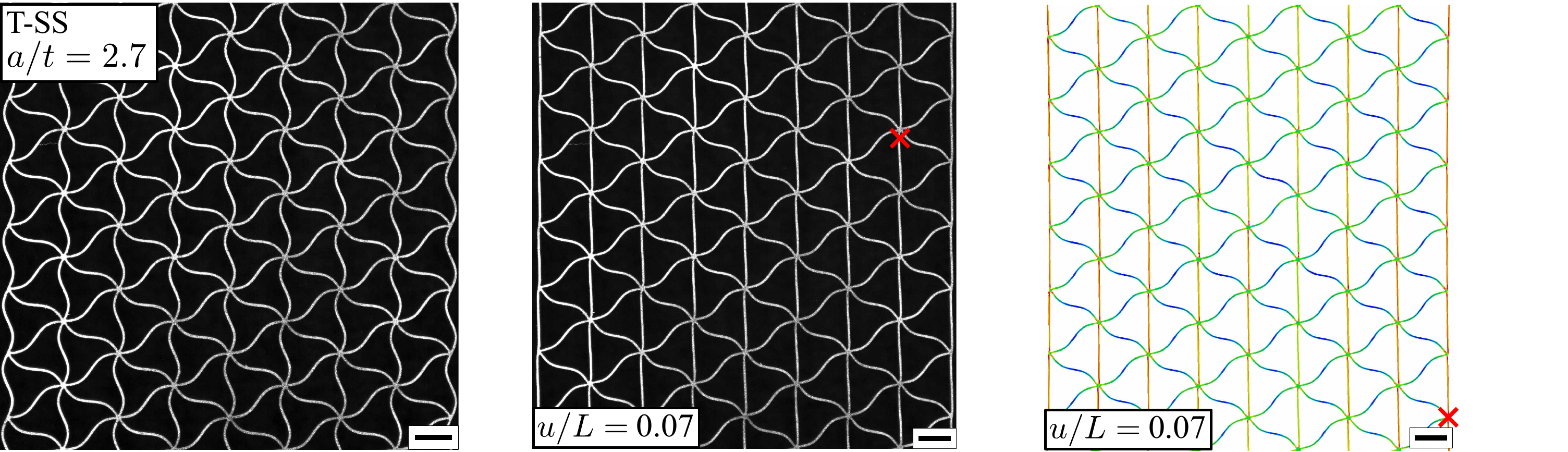}\label{fig:failureSequence_tria_b}}\\[\baselineskip]
    \sidesubfloat[]{\includegraphics[width=0.99\textwidth]{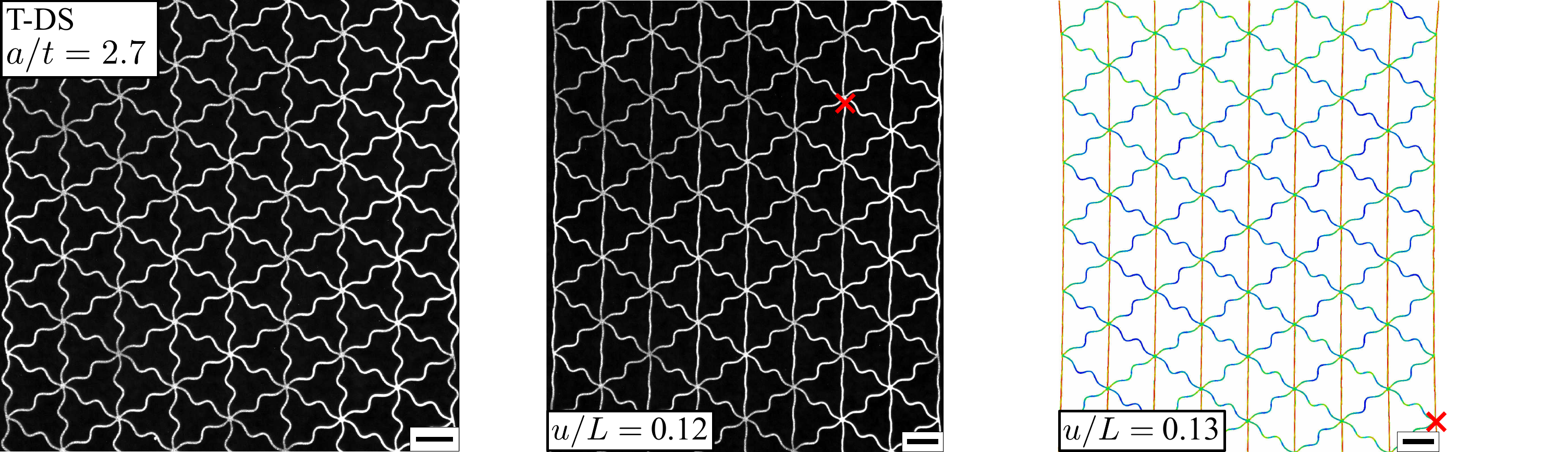}\label{fig:failureSequence_tria_c}}\\[\baselineskip]
    \caption{Experimental observations and FE predictions of the deformed
      triangular lattices at the peak load. The experimental images of
      the undeformed lattices are also
      included. \subref{fig:failureSequence_tria_a} Lattice with
      straight (S) struts; \subref{fig:failureSequence_tria_b}
      sinusoidal shaped struts with $a/t=2.7$ and
      \subref{fig:failureSequence_tria_c} decaying sinusoidal struts
      with $a/t=2.7$. The locations of first strut failure are marked
      on the experimental and FE images with the FE images showing
      contours of the Von-Mises stress. The scale bar is of length
      \SI{15}{mm}.}
  \label{fig:failureSequence_tria}
\end{figure}

\begin{figure}[htbp]
  \centering
    \setlength{\labelsep}{0.25cm}
    \sidesubfloat[]{\includegraphics[width=0.99\textwidth]{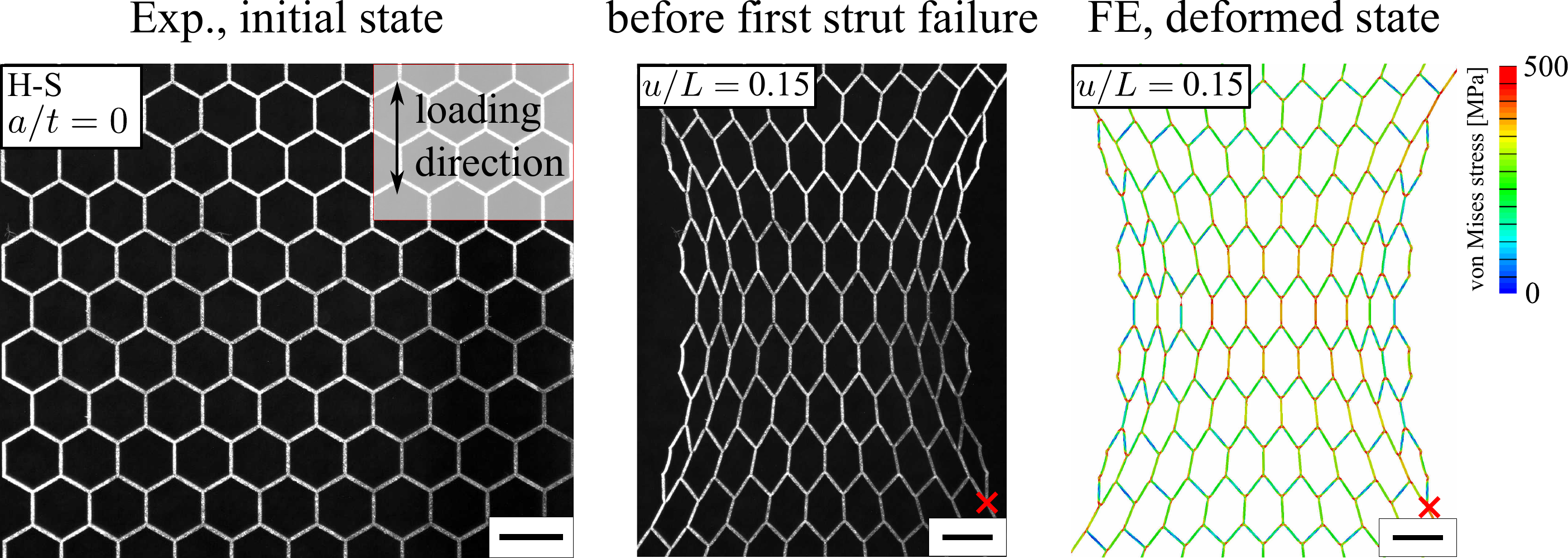}\label{fig:failureSequence_hex_a}}\\[\baselineskip]
    \sidesubfloat[]{\includegraphics[width=0.99\textwidth]{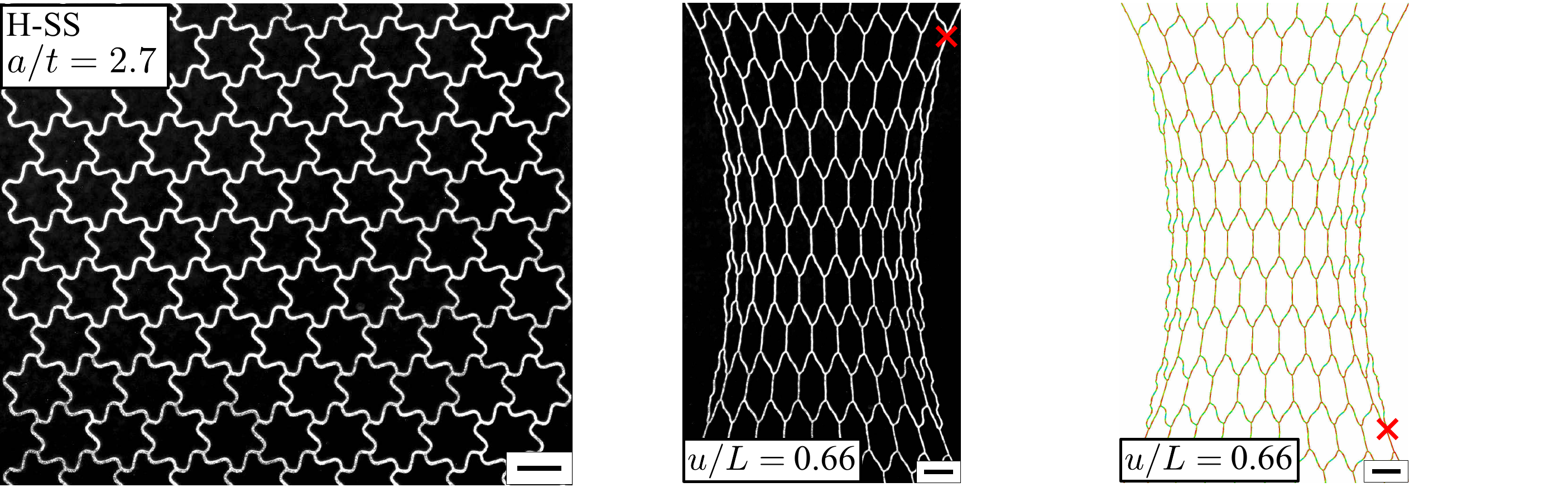}\label{fig:failureSequence_hex_b}}\\[\baselineskip]
    \sidesubfloat[]{\includegraphics[width=0.99\textwidth]{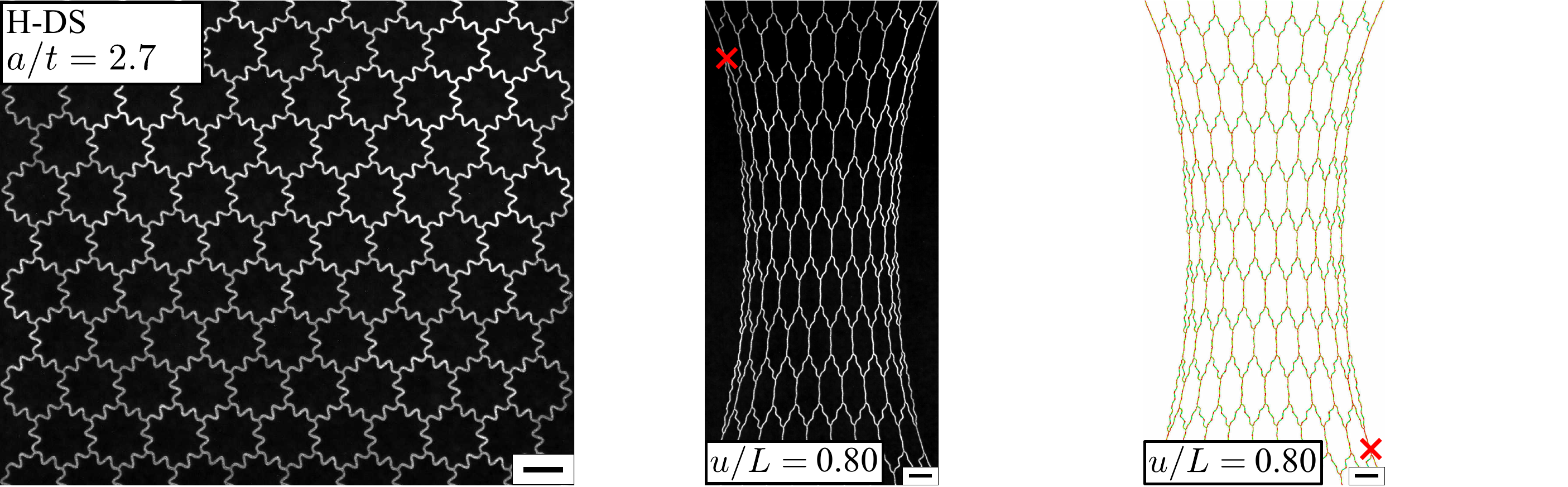}\label{fig:failureSequence_hex_c}}\\[\baselineskip]
    \caption{Experimental observations and FE predictions of the deformed hexagonal
      lattices at the peak load. The experimental images of the
      undeformed lattices are also
      included. \subref{fig:failureSequence_hex_a} Lattice with
      straight (S) struts; \subref{fig:failureSequence_hex_b}
      sinusoidal shaped struts with $a/t=2.7$ and
      \subref{fig:failureSequence_hex_c} decaying sinusoidal struts
      with $a/t=2.7$. The locations of first strut failure are marked
      on the experimental and FE images with the FE images showing
      contours of the Von-Mises stress. The scale bar is of length
      \SI{15}{mm}.}
  \label{fig:failureSequence_hex}
\end{figure}

The measured nominal stress versus nominal strain responses until
first strut failure are in
\cref{fig:stressStrain_constRelD_combined2}, with the peak load
$P_\mathrm{f}$ occurring at first strut failure; the macroscopic
ductility is defined as $\varepsilon_\mathrm{f}\equiv u_\mathrm{f}/L$,
where $u_\mathrm{f}$ is the displacement corresponding to the load
$P_\mathrm{f}$. The measured mean undercut depths $e/t$ for each
specimen are reported in \cref{tab:geometry_constRelD}. In
\cref{fig:stressStrain_constRelD_combined2} measurements are included
for both triangular (T) and hexagonal (H) lattices with straight (S)
struts and wavy struts ($0<a/t\le 2.7$), with ``SS'' and ``DS''
referring to sinusoidal and \ds sinusoidal shaped struts,
respectively. The choice $a/t=0$ refers to straight (S) struts and this
limiting case is included in
\cref{fig:stressStrain_constRelD_combined2}. The corresponding FE
simulations are included for all cases in
\cref{fig:stressStrain_constRelD_combined2} and are terminated at the
attainment of peak load: similar to the measurements, necking of a
strut was detected in the FE simulations at peak load. These FE
simulations assumed the measured value of $e/t$ for each specimen listed
in \cref{tab:geometry_constRelD}.
\begin{enumerate}[(i)]
\item Consider lattices comprising straight struts, with images of
the undeformed and deformed triangular and hexagonal lattices shown in
Figs. \ref{fig:failureSequence_tria} and
\ref{fig:failureSequence_hex}, respectively, along with corresponding
FE predictions at peak load; contours of von-Mises stress are shown on
the FE images. These lattices behave according to the regimes defined
in \cite{Tankasala2017}. 

The stretch-dominated triangular lattice has
a response that can be divided into 3 regimes. In regime~I the
vertical struts that are aligned with the loading direction undergo
elastic stretching while the inclined struts rotate. In regime~II, the
vertical struts undergo plastic stretching with first failure
occurring in these struts due to necking. Since failure occurs in
regime~II, these measurements do not display a regime~III wherein the
inclined struts rotate to align with the loading
direction~\cite{Tankasala2017}. Failure occurs at a relatively low
macroscopic strain level and deformation of the lattice at first
failure is barely visible as seen in \cref{fig:failureSequence_tria}
(the location of the first strut failure is marked in
\cref{fig:failureSequence_tria}). 

For the bending-dominated hexagonal
lattices with straight struts (H-S), regimes~I and II are
characterised by elastic and plastic bending, respectively, of the
struts of the lattice. This bending-dominated response implies that
the hexagonal lattice has a high initial compliance. Rotation of the
inclined struts in regime~II aligns them with the loading direction
and thereafter the response enters into regime~III. In this regime,
the struts of the bending-governed hexagonal lattice stretch and then
fail by necking. While the ductility of the lattice struts is
approximately the same for the triangular and hexagonal lattices the
macroscopic strain associated with bending and rotation of the lattice
struts endows the hexagonal lattice with a greater ductility than that
of the triangular lattice ($\varepsilon_\mathrm{f}\approx 3\,\%$ and
15\,\% for the triangular and hexagonal lattices,
respectively). Moreover, unlike the triangular lattice
(\cref{fig:failureSequence_tria}), tensile stretching of the
hexagonal lattice involves significant transverse contraction as seen
in the deformed images in \cref{fig:failureSequence_hex}. This
contraction is inhibited by the gripping of the lattice at the top and
bottom of the specimens and subsequently we shall show that this
results in enhanced stresses in the struts at the edge of the
hexagonal lattice specimens. As a consequence, the first strut that
fails is at the edge of the specimen (marked in
\cref{fig:failureSequence_hex}).

\item Consider the lattices with wavy struts with the nominal stress
  versus strain responses given in
  \cref{fig:stressStrain_constRelD_combined2}, and the corresponding
  images of the deformed lattices given in
  Figs.~\ref{fig:failureSequence_tria} and
  \ref{fig:failureSequence_hex}. While the peak load $P_\mathrm{f}$
  decreases only mildly with increasing waviness amplitude $a/t$, the
  ductility $\varepsilon_\mathrm{f}$ of both the triangular and
  hexagonal lattices increase substantially with increasing $a/t$.
  Note that the drop in ductility for $a/t=0.7$ and 1.3 of T-SS
  lattices in \cref{fig:stressStrain_constRelD_tria_sine_at} is due to
  the increased undercut depth $e/t$ as reported in
  \cref{tab:geometry_constRelD}. For a given lattice topology and
  waviness amplitude $a/t$, struts of \ds sinusoidal shape result in
  higher ductility than struts of sinusoidal shape in line with the
  single strut results of \cref{sec:single_strut_resp}. Also, the
  bending-dominated hexagonal lattices have a higher ductility than
  the stretching-dominated triangular lattices. The images in
  Figs.~\ref{fig:failureSequence_tria} and
  \ref{fig:failureSequence_hex} show that, at the instant of first
  strut failure (location marked in both figures), all struts of the
  hexagonal lattice have lost their waviness by axial stretch while
  the inclined struts of the triangular lattice retain significant
  waviness. Failure of the triangular lattices occurs after the
  waviness in the vertical struts has been eliminated. On the other
  hand, bending-dominated deformation of the hexagonal lattices
  implies significant scissoring of the struts; strut stretching,
  required to neck the struts, initiates only after all struts have
  aligned with the loading direction and waviness has been eliminated.

The FE predictions of the tensile responses of the lattices are
included in \cref{fig:stressStrain_constRelD_combined2} while
predictions of the deformed configurations at peak load are presented
in Figs.~\ref{fig:failureSequence_tria} and
\ref{fig:failureSequence_hex}. Recall that the $e/t$ value for each
specimen is different and is listed in
\cref{tab:geometry_constRelD}. Upon assuming the appropriately chosen
value of $e/t$, excellent agreement is observed between the FE
predictions and measurements including the deformed lattice
shapes. However, we emphasise that $e/t$ has a strong influence and
this is mapped out in detail in \cref{sec:design_maps}.

\end{enumerate}

\subsection{Effect of finite specimen size}
\label{sec:specimen_sizes}
The above results suggest that the constraint imposed by the gripping
of the specimens results in the development of high stresses in struts
along the side edges of the specimens
(\cref{fig:failureSequence_hex}). This is particularly pronounced for
the hexagonal lattices as they have a high value of plastic Poisson's
ratio. Here we use FE simulations to investigate the effect of finite
specimen geometry upon the tensile responses of the lattices by
contrasting specimen predictions with those of corresponding infinite
lattices. The infinite lattices were simulated by considering a
representative volume element (RVE) comprising a single unit cell, and
uniaxial tension was simulated by imposing periodic boundary
conditions on this RVE. All struts in the RVE had an undercut of
normalised depth $e/t=0.1$ and, in order to make a fair comparison, we
also report FE predictions of the tensile responses of lattice
specimens (of identical geometry to those considered above) but with
all struts in the lattice having an undercut of depth
$e/t=0.1$. Predictions are given up to the onset of necking in any
strut within the lattice; this point also corresponds to the
attainment of peak macroscopic load in the predictions.

\begin{figure}[htbp]
  \centering
  \setlength{\labelsep}{-0.95cm}
  \sidesubfloat[]{\includegraphics[width=0.50\textwidth]{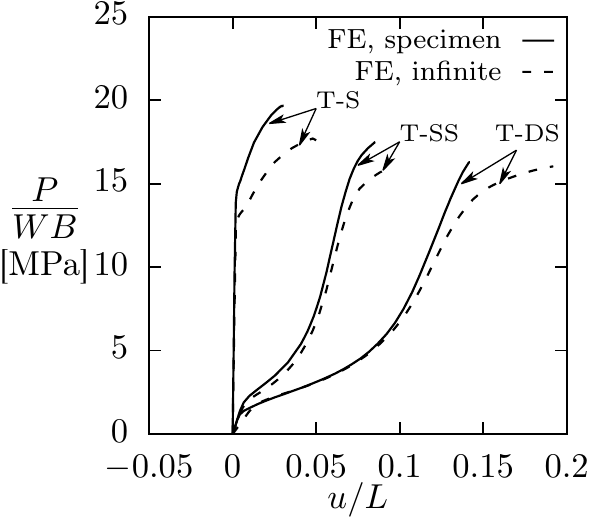}\label{fig:stressStrain_constRelD_combined_FE_T}}\hfill
  \sidesubfloat[]{\includegraphics[width=0.49\textwidth]{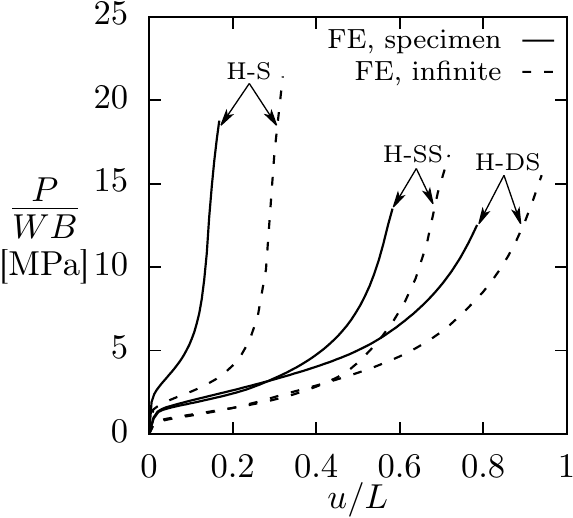}\label{fig:stressStrain_constRelD_combined_FE_H}}
  \caption{FE predictions of the stress versus strain responses of the
    finite lattice specimens and infinite periodic lattices with
    straight and wavy struts ($a/t=2.7$). An undercut of depth
    $e/t=0.1$ was used in all
    calculations. \subref{fig:stressStrain_constRelD_combined_FE_T}
    Triangular lattices and
    \subref{fig:stressStrain_constRelD_combined_FE_H} hexagonal
    lattices.}
  \label{fig:stressStrain_constRelD_combined}
\end{figure}

Consider the triangular lattice responses shown in
\cref{fig:stressStrain_constRelD_combined_FE_T} for straight,
sinusoidal and \ds sinusoidal strut shapes ($a/t=2.7$ for wavy
struts). The difference in responses is small for the finite lattice
specimen and infinite lattice although we observe that the infinite
lattice has a slightly higher ductility due to a more compliant
response just prior to peak load. These results can be contrasted to
the corresponding hexagonal lattice predictions included in
\cref{fig:stressStrain_constRelD_combined_FE_H}. While the peak
strengths of the infinite and finite hexagonal lattices are
approximately equal, the ductility of the infinite lattices is
significantly higher in all cases (i.\,e. straight, sinusoidal and \ds
sinusoidal strut shapes). This can be understood from the deformed
lattice specimen images in \cref{fig:failureSequence_hex}. The
constraint of the grips limits the degree of plastic Poisson
contraction of the hexagonal lattices, thereby straightening the
struts at the specimen sides at smaller applied values of $u/L$. A
consequence of this straightening is a build-up of tensile stress in
struts at the specimen sides which in turn results in increasing
hardening of the tensile response and premature necking of the edge
struts.

\section{Design maps}
\label{sec:design_maps}
There is a strong interplay between the waviness amplitude $a/t$ and
undercut depth $e/t$ that sets the peak strength $P_\mathrm{f}$ and
ductility $\varepsilon_\mathrm{f}$. Here we employ FE calculations to
explore this interplay for the lattice specimens of
\cref{sec:tensile_response_lattices} with the aim of providing
guidance on the design of wavy lattices to achieve a specified
strength and ductility.

\begin{figure}[htbp]
  \centering
  \setlength{\labelsep}{-0.1cm}
  \sidesubfloat[]{\includegraphics[width=0.44\textwidth]{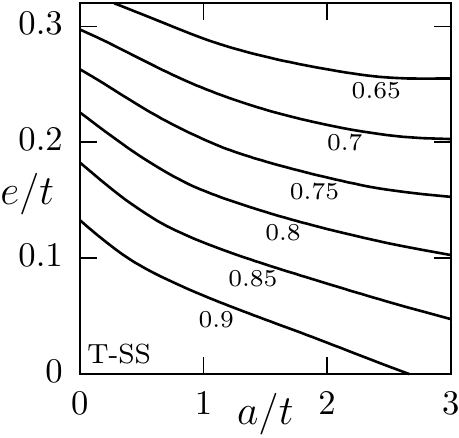}\label{fig:contourStrength_T_SS}}\hfill
  \sidesubfloat[]{\includegraphics[width=0.44\textwidth]{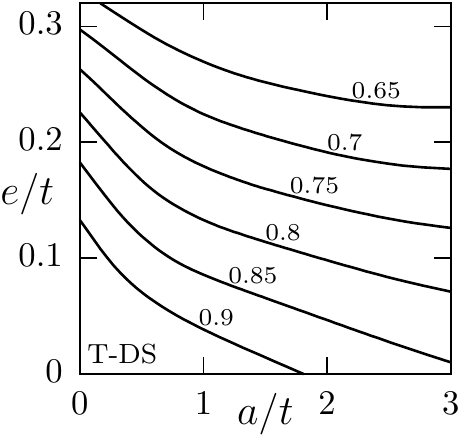}\label{fig:contourStrength_T_DS}}\\[\baselineskip]
  \sidesubfloat[]{\includegraphics[width=0.44\textwidth]{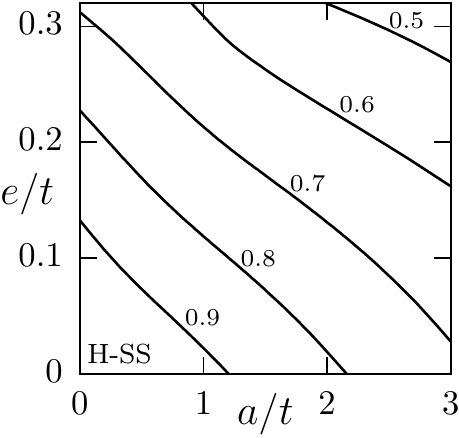}\label{fig:contourStrength_H_SS}}\hfill
  \sidesubfloat[]{\includegraphics[width=0.44\textwidth]{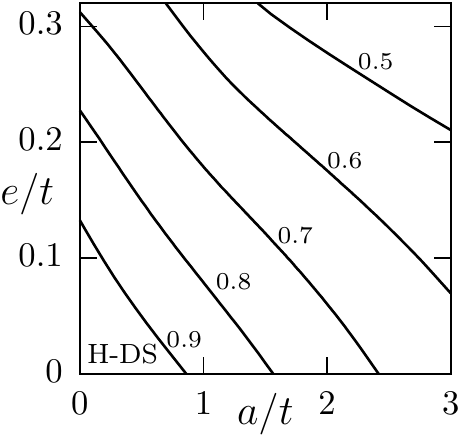}\label{fig:contourStrength_H_DS}}
  \caption{Design map of the predicted knock-down
    $P_\mathrm{f}/P_\mathrm{f}^0$ with axes of normalised undercut
    depth $e/t$ and strut waviness amplitude $a/t$ for triangular
    lattices with \subref{fig:contourStrength_T_SS} sinusoidal (SS)
    and \subref{fig:contourStrength_T_DS} \ds sinusoidal triangular
    (DS) shaped struts. The corresponding hexagonal lattices predictions for the
    \subref{fig:contourStrength_H_SS} sinusoidal (SS) and
    \subref{fig:contourStrength_H_DS} decaying sinusoidal (DS) shaped struts.}
  \label{fig:contourStrength}
\end{figure}

FE predictions of contours of normalised tensile failure strength
$P_\mathrm{f}/P_\mathrm{f}^0$ of the triangular lattice with
sinusoidal and \ds sinusoidal shaped struts, are given in
\cref{fig:contourStrength_T_SS,fig:contourStrength_T_DS},
respectively, in the form of a map with axes of $a/t$ and $e/t$.  Each
contour plot is generated by 20 FE simulation. Here, $P_\mathrm{f}$ is
the failure strength of the lattice for the given choice $(a/t,e/t)$
while $P_\mathrm{f}^0$ is the reference strength of the perfect
lattice with $e/t=a/t=0$. The corresponding predictions for the
hexagonal lattice with sinusoidal and \ds sinusoidal shaped struts are
given in \cref{fig:contourStrength_H_SS,fig:contourStrength_H_DS},
respectively. In all cases, the knockdown in strength, as
parameterised by $P_\mathrm{f}/P_\mathrm{f}^0$, increases with
increasing $a/t$ and $e/t$; the waviness amplitude $a/t$ has a larger
effect on the hexagonal lattice while the undercut depth $e/t$ plays a
more dominant role for the triangular lattices. This is evident from
the orientation of the $P_\mathrm{f}/P_\mathrm{f}^0$ contours.

\begin{figure}[htbp]
  \centering
  \setlength{\labelsep}{-0.1cm}
  \sidesubfloat[]{\includegraphics[width=0.44\textwidth]{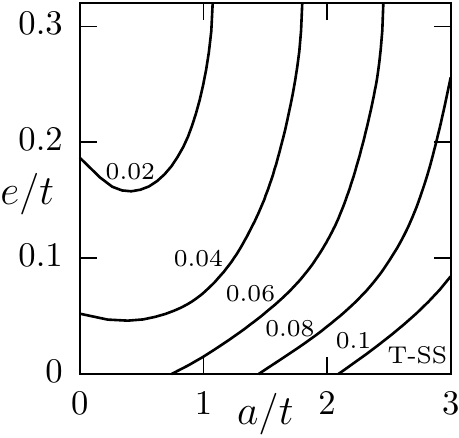}\label{fig:contourDuctility_T_SS}}\hfill
  \sidesubfloat[]{\includegraphics[width=0.44\textwidth]{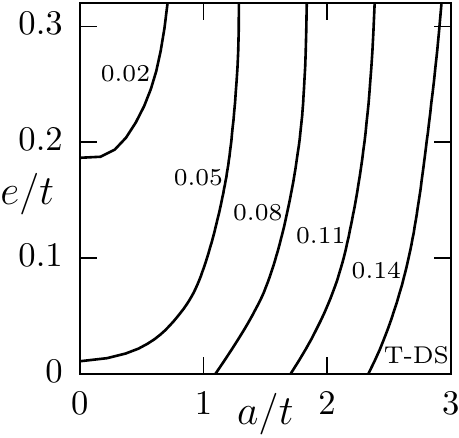}\label{fig:contourDuctility_T_DS}}\\[\baselineskip]
  \sidesubfloat[]{\includegraphics[width=0.44\textwidth]{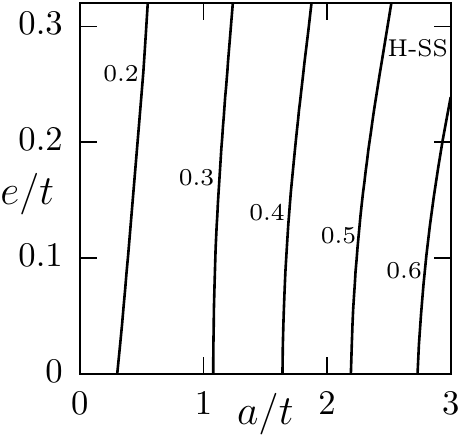}\label{fig:contourDuctility_H_SS}}\hfill
  \sidesubfloat[]{\includegraphics[width=0.44\textwidth]{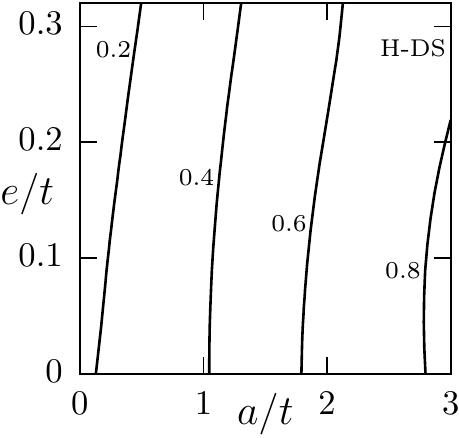}\label{fig:contourDuctility_H_DS}}
  \caption{Design map of the predicted ductility $u_\mathrm{f}/L$
    with axes of normalised undercut depth $e/t$ and strut waviness
    amplitude $a/t$ for triangular lattices with
    \subref{fig:contourDuctility_T_SS} sinusoidal (SS) and
    \subref{fig:contourDuctility_T_DS} \ds sinusoidal triangular (DS)
    shaped struts. The corresponding hexagonal lattices predictions for the
    \subref{fig:contourDuctility_H_SS} sinusoidal (SS) and
    \subref{fig:contourDuctility_H_DS} decaying sinusoidal (DS) shaped
    struts.}
  \label{fig:contourDuctility}
\end{figure}

Next, consider the effect of $a/t$ and $e/t$ upon ductility. FE
predictions of contours of the ductility $\varepsilon_\mathrm{f}$
generated by 20 FE simulations are plotted in
\cref{fig:contourDuctility} on a design map with axes $a/t$ and
$e/t$. The contours of $\varepsilon_\mathrm{f}$ are nearly vertical
for the hexagonal lattices
(\cref{fig:contourDuctility_H_SS,fig:contourDuctility_H_DS} for
lattices with sinusoidal and \ds sinusoidal shaped struts,
respectively) indicating that the presence of the undercut does not
significantly degrade ductility. Now consider the contour plots of
\cref{fig:contourDuctility_T_SS,fig:contourDuctility_T_DS} for
triangular lattices with sinusoidal and \ds sinusoidal shaped struts,
respectively. The contours of $\varepsilon_\mathrm{f}$ become more
horizontal at low values of $a/t$ suggesting that the ductility of
triangular lattices with nearly straight struts is largely governed by
the undercut depth. Consistent with the findings of the experiments
and FE simulations reported in \cref{sec:tensile_response_lattices},
the maps in \cref{fig:contourDuctility} show that hexagonal lattices
have a higher ductility than triangular lattices. Moreover, for a
given lattice topology, lattices with a \ds sinusoidal shaped struts
have a higher ductility then the corresponding lattices with
sinusoidal shaped struts.

\section{Concluding remarks}
\label{sec:concluding_remarks}
Our study has explored, by a combination of measurements and finite element (FE)
simulations, the sensitivity of tensile response of bending-dominated
hexagonal lattices and stretching-dominated triangular lattices to
strut shape. Lattices of relative density $\bar\rho=0.1$ were
manufactured by water-jet cutting of \SI{3}{mm} thick low-carbon steel
sheets. Two strut shapes (sinusoidal and \ds sinusoidal) of varying
amplitude were investigated, alongside the role of manufacturing
defects such as undercuts in the struts near the lattice
nodes. Excellent agreement between the measurements and FE simulations
allowed us to proceed to employ FE simulations to develop design maps.

An increased strut waviness greatly enhances the ductility of both
types of lattice but has a smaller effect upon the peak tensile
strength. Moreover, for a given waviness amplitude, the lattices with
\ds sinusoidal shaped struts have the highest ductility. The increase
in the ductility of stretch-dominated triangular lattices with
increased waviness is mainly due to the fact that waviness in the
vertical struts needs to be ironed-out prior to them undergoing
stretching and then necking. On the other hand, the large rotation of
the struts in the bending-dominated hexagonal lattices implies that
waviness in all struts needs to be eliminated prior to strut
necking. Thus, the ductility enhancement due to waviness is higher in
the hexagonal lattices. Imperfections such as undercuts in the lattice
strut have a larger effect on the ultimate tensile strength than
ductility, and this is demonstrated over a large parameter range via
design maps as developed by FE calculations.

\begin{acknowledgment}
The authors gratefully acknowledge the financial support from the
European Research Council (ERC) under the European Union's Horizon
2020 research and innovation program, grant GA669764, MULTILAT.
\end{acknowledgment}

\bibliographystyle{asmems4}
\bibliography{Literature}

\clearpage
\listoftables
\listoffigures

\clearpage
\section*{Tables}

\begin{table}[htbp]
  \begin{tabular}{ l | l | c | c | c | c | c}
    lattice geometry & strut shape & $a/t$ & $t/\ell$ &
                                                        $\ell_\mathrm{s}/\ell$
    & $t\ell_\mathrm{s}/\ell^2$ & $e/t$\\
    \hline
    triangular & straight (T-S) & 0 & 0.030 & 1.00 & 0.03 & 0.2\\
    \cline{2-7}
                     & & 0.7 & 0.030 & 1.01 & 0.03 & 0.3\\
                     & sine (T-SS) & 1.3 & 0.029 & 1.02 & 0.03 & 0.3\\
                     &  & 2.7 & 0.028 & 1.05 & 0.03 & 0.3\\
    \cline{2-7}
                     &  & 0.7 & 0.029 & 1.01 & 0.03 & 0.3\\
                     & \ds sine (T-DS) & 1.3 & 0.028 & 1.03 & 0.03 & 0.2\\
                     &  & 2.7 & 0.027 & 1.11 & 0.03 & 0.3\\
    \hline
    hexagonal & straight (H-S) & 0 & 0.089 & 1.00 & 0.09 & 0.2\\
    \cline{2-7}
                     &  & 0.7 & 0.086 & 1.03 & 0.09 & 0.3\\
                     & sine (H-SS) & 1.3 & 0.080 & 1.11 & 0.09 & 0.3\\
                     &  & 2.7 & 0.069 & 1.28 & 0.09 & 0.2\\
    \cline{2-7}
                     &  & 0.7 & 0.083 & 1.07 & 0.09 & 0.2\\
                     & \ds sine (H-DS) & 1.3 & 0.075 & 1.19 & 0.09 & 0.2\\
                     &  & 2.7 & 0.062 & 1.43 & 0.09 & 0.2\\

  \end{tabular}
  \caption{Geometric parameters of the hexagonal and triangular
    lattices investigated in this study. All lattices had a relative
    density $\bar\rho=0.1$ with the geometric parameters $a$, $t$,
    $\ell$, and $\ell_ \mathrm{s}$ as specified in the CAD input. The
    undercuts of depth $e$ were a consequence of the manufacturing
    process and we tabulate here the mean value of $e/t$ measured over
    all struts for each specimen.}
  \label{tab:geometry_constRelD}
\end{table}

\clearpage
\section*{Figures}
\floatsetup[figure]{subcapbesideposition=top}

\thispagestyle{empty}

\end{document}